\documentclass[aps,prf,floatfix]{revtex4-2}

\usepackage{graphicx}
\usepackage{epstopdf, epsfig}

\usepackage{color}
\definecolor{colortodo}{RGB}{255,0,0}
\newcommand{\red}[1]{{\color{colortodo}#1}}

\definecolor{colortodo2}{RGB}{0,255,0}

\newcommand{\rev}[1]{{{#1}}}
\newcommand\revDEL[1]{}

\usepackage{upgreek}
\usepackage{natbib}
\usepackage{ulem}
\usepackage{hyperref}
\hypersetup{
colorlinks=true, 
citecolor=magenta,
breaklinks=true, 
urlcolor= blue, 
linkcolor= blue, 
bookmarksopen=true, 
}

\usepackage[utf8]{inputenc}   
\usepackage[T1]{fontenc}      

\usepackage{lmodern} 					

\usepackage{amsfonts}
\usepackage{amssymb}
\usepackage{amsmath}

\usepackage{epstopdf}
\usepackage{pifont}
\usepackage{color}
\usepackage{xcolor}
\usepackage{wasysym}
\usepackage{exscale}
\usepackage[caption=false]{subfig}

\usepackage{morefloats}

\begin{document}

\title{Experimental study of three-dimensional turbulence under a free surface}

\author{Timoth\'{e}e Jamin}
\email[E-mail: ]{research@timotheejamin.com}
\author{Michael Berhanu}
\author{Eric Falcon}
\affiliation{Universit\'{e} Paris Cit\'{e}, MSC, UMR 7057 CNRS, F-75013 Paris, France}

\date{\today}

\begin{abstract}
In many environmental flows, an air-water free surface interacts with a turbulent flow in the water phase. To mimic this situation, we propose an original experimental setup, based on the Randomly Actuated Synthetic Jet Array (RASJA) forcing used to study turbulence with almost no mean flow. By using a central pump connected to jets, we generate a turbulent flow of tunable intensity, with good isotropy and horizontal homogeneity, up to a maximal turbulent Reynolds number of $8800$. {The aim here is to characterize how turbulence is modified by the presence of a free surface. We show that the free surface affects turbulence for fluid depths smaller than the integral length.} We report that the turbulent fluctuations become strongly anisotropic when approaching the free surface. The vertical velocity fluctuations decrease close to the surface whereas the horizontal ones increase, as reported in previous theoretical predictions and numerical observations. We also observe a strong enhancement of the amplitude of the temporal and spatial power spectra of the horizontal velocity at large scales, showing the strengthening of these velocity fluctuations near the free surface. {Besides, these behaviors are studied for various levels of turbulence.}
\end{abstract}

\maketitle

\section{Introduction}

{In many environmental flows, such as ocean currents, rivers, or streams, the turbulence in the water interacts with the free surface at the interface between ocean and atmosphere.  Gas, heat, and mass exchanges~\cite{ThorpeReview1995} can be considerably increased at the interface in {the} presence of turbulence and this phenomenon warranted several studies in the field~\cite{VeronJPO2009}, with controlled experiments~\cite{HerlinaJFM2008,VarianoJFM2013} or using numerical simulations~\cite{HerlinaJFM2019}. Brocchini and Peregrine~\cite{BrocchiniJFM2001} propose a qualitative description of a turbulent free surface. To cope with {the} complexity of mutual interactions between the turbulent flow and the deformable interface, a first step consists in understanding the effect of a free surface on the hydrodynamic turbulence. We can distinguish two standard cases depending on whether turbulence is accompanied by a mean flow. In the first, several studies investigated open channel flows numerically~\cite{pan1995,ShenJFM2001,YoshimuraHR2019} and experimentally~\cite{kumar1998experimental}. The degree of turbulence can also be enhanced using an active grid~\cite{SavelsbergPRL2008,SavelsbergJFM2009} or a rough bed in shallow water~\cite{DolcettiPOF2016}. For deep-water fluids, the ratio of velocity fluctuations to the mean flow remains below $10\,\%$, even using an active grid.}

{In this study, we focus on {the case with no} mean flow. Several numerical simulations} have investigated homogeneous and isotropic turbulence in the vicinity of a free-slip top interface in a decaying regime or with sustained forcing~\cite{Perot1995,CalmetMagnaudetJFM2003, CampagnePOF2009, HerlinaJFM2014, FloresJFM2017, HerlinaJFM2019}. To approach natural and experimental configurations, a few studies have also investigated the case of a free surface contaminated by surfactants~\cite{ShenJFM2004,BodartJturb2010,WissinkJFM2017}. 
Two main results can be summarized from the previous studies. First, near the free surface, the vertical velocity fluctuations are significantly reduced whereas the horizontal ones are enhanced. The rapid distortion theory (RDT) can explain this effect~\cite{HuntJFM1978,TeixeiraJFM2000,MagnaudetJFM2003}.
Secondly, for a high enough Reynolds number, the power spectrum of the horizontal velocities follows the power law in $k^{-5/3}$ predicted by Kolmogorov for isotropic homogeneous turbulence~\cite{TennekesBook,Pope} even close to the free surface despite the flow anisotropy. Some experimental studies report this power law close to the free surface from field observations on the surface of a river~\cite{chickadel2011infrared,TalkeJGRO2013} and in laboratory experiments with a grid-stirred tank~\cite{BrumleyJFM1987,HerlinaJFM2008}. This observation may be explained by an analogy with stratified turbulence~\cite{FloresJFM2017}.

Most of the experiments investigating the interaction between the turbulence with low mean flow and a free surface use an oscillating grid to generate turbulence~\cite{thompson1975mixing,BrumleyJFM1987,BadulinOcean1988,McKennaIJHM2004, chiapponi2012experimental}, where a horizontal grid plunged in the tank below the free surface is vertically oscillated. {According to Variano and Cowen~\cite{VarianoJFM2008}}, these devices have experimental repeatability issues~\cite{mckenna2000free} and the level of turbulence remains moderate: typically the root mean square (rms) velocity is below or equal to $5\,$cm s$^{-1}$ for a meter scale experiment. {The resulting flows} display also large-scale correlations due to the grid oscillation. Large-scale perturbations of the free surface would be also expected by a turbulent flow produced by the motion of impellers like for the von K\'arm\'an closed flows~\cite{douady1991direct,fauve1993pressure,RaveletJFM2008} or by the erratic motion of small magnetically-driven {particles~\cite{Cazaubiel2021,Gorce2022}}. However, these methods with inertial forcing have not been used for now to study turbulence in the vicinity of a free surface. {The free-surface can also be forced by anisotropic turbulence using  {magnetohydrodynamic} forcing in liquid metals~\cite{GutierrezPOF2016}.}

Finally, {an efficient} method to investigate turbulence near a free surface in the laboratory consists in using a water tank randomly stirred by synthetic jets placed at the bottom of the tank. This configuration introduced and characterized by E. Variano, E.~Bodenschatz, and E. Cowen~\cite{VarianoEF2004,VarianoJFM2008} displays a turbulent flow featuring high-Reynolds-number, large-scale isotropy, homogeneity and a ratio between mean and fluctuating flow less than 10\,\%. Like in the previous studies using an oscillating grid~\cite{BrumleyJFM1987,chiapponi2012experimental}, the decay of vertical fluctuations when approaching the free surface is in qualitative agreement with the predictions of the RDT~\cite{HuntJFM1978,MagnaudetJFM2003,CalmetMagnaudetJFM2003}. {As summarized in a recent review~\cite{Nezami2023}}, random jet arrays have become a popular method to generate homogeneous isotropic turbulence with {low} mean flow in air and water using various configurations of the array~\cite{HwangEF2004,BellaniEF2014,CarterEF2016,JohnsonJFM2017,EstebanJFM2019} and different driving algorithms firing the jets~\cite{PerezAvaradoEF2016}. However, for experiments in water, jets are produced by numerous (from $60$ to $128$) inexpensive immersed small pumps, which are set on or off according to a firing pattern. {As the flow rate of these pumps is fixed by their working point, the global intensity of the turbulence cannot be varied easily without changing other physical quantities}. We note also that turbulence in the presence of a free surface has been investigated in a jet-agitated vessel~\cite{George1994,Minel1995} to measure gas-liquid transfers. One hundred vertical nozzles at the bottom are connected to a large pump. {Yet}, with jets running continuously, secondary motions very sensitive to initial conditions induce the presence of a non-negligible mean flow~\cite{Minel1995}.

{Here, {we study} experimentally the interaction between hydrodynamic turbulence and a deformable free surface. Based on the Randomly Actuated Synthetic Jet Array (RASJA) forcing method~\cite{VarianoEF2004,VarianoJFM2008}, we propose an original setup that allows the turbulence level to be varied easily}. A large pump with an adjustable and regulated flow rate is connected with an array of 64 solenoid valves controlled independently. By varying the global flow rate, a large range of levels of turbulence can be obtained in a water tank of $40\times40\times 75\,$cm while keeping a good level of homogeneity and isotropy for an appropriate choice of open jet duration and firing pattern. Using this device, we characterize the near-surface turbulence and fruitfully complete existing works. The study of free-surface deformation induced by turbulence~\cite{GuoJFM2010} will be the object of subsequent work. The paper is organized as follows. Section~\ref{Experimentalsetup} describes the experimental setup and the methods to generate nearly homogeneous and isotropic turbulence with a small mean flow in a water tank. In Sec. \ref{Homogeneousturbulence}, we characterize the turbulence far from the free surface for the best choice of the jet pattern. Then, Sect.~\ref{sec:nearFS} investigates the modification of the turbulence when the free surface is approached. We give our conclusions and perspectives in Sec.~\ref{Conclusion}.

\section{Experimental setup}
\label{Experimentalsetup}

\subsection{A randomly actuated synthetic jet array with a tunable flow-rate}


{Based on the RASJA forcing method~\cite{VarianoJFM2008}, we propose here an experimental setup where the level of turbulence can be varied independently of the injection geometry.} The jets are supplied by only one central centrifugal pump with a high flow rate and each jet is individually driven by a solenoid valve which turns the jet on or off by being respectively open or closed.  Due to the central high flow rate of the pump, the turbulence intensity can be varied over a large range of Reynolds numbers. 
This feature is of primary importance for our setup as the final objective consists in characterizing different regimes of surface deformation as a function of the subsurface Reynolds number. {A similar configuration had been used by Variano \textit{et al.} as a prototype with 9 jets \cite{VarianoEF2004} and by Delbos \textit{et al.} \cite{delbos2009homogeneous}, but using 4 solenoid valves only to control 64 jets. None of these two studies fully characterized the obtained turbulence.}

\begin{figure}
\centering
\includegraphics[width=\textwidth]{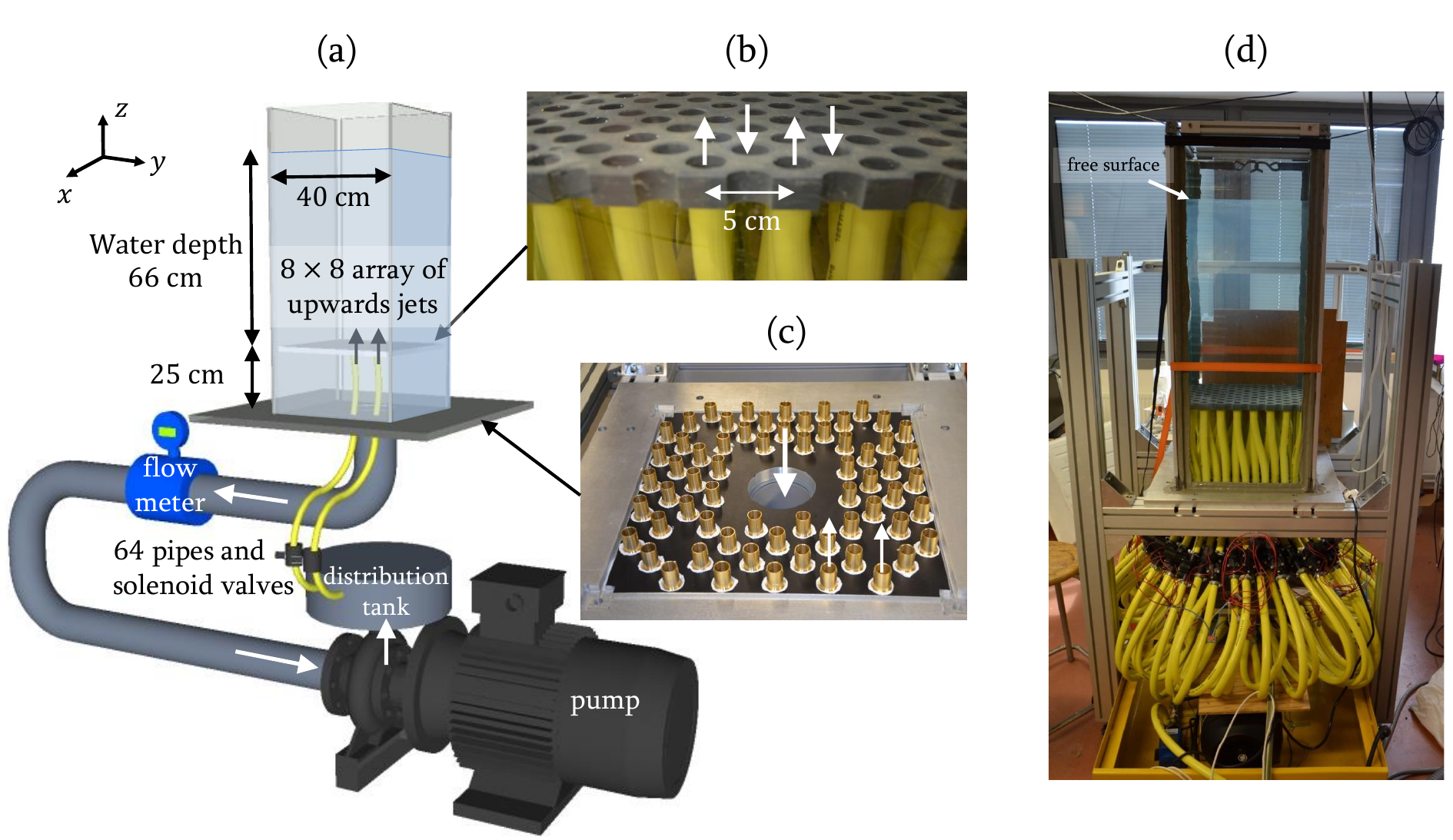}
\caption{ (a) General schematic diagram of the experimental setup used to produce turbulence. The turbulent flow is generated in a square glass tank of side $40\,$cm filled with water to a  depth of $66\,$cm. (b) Close-up view of the 64-jet grid (up arrows) with a 5-cm spacing placed 25\,cm above the tank bottom. Down arrows correspond to suction holes. (c) View of the tank bottom before connecting the pipes. Jets are arranged around a central suction hole. (d) Global view of the experimental setup.}
\label{fig:setup_turbulence}
\end{figure}

In our setup, the turbulence is produced with 64 jets in a glass tank, which is $40\times 40~$cm$^2$ wide and 100~cm high [see Fig.~\ref{fig:setup_turbulence}\,(a)]. The jets are placed 25\,cm above the tank bottom and are firing upwards. They are arranged in a $8\times 8$ array with a spacing of 5\,cm. The flow is produced by a unique centrifugal pump. The outlet port of the pump is facing up and is connected to a cylindrical redistribution chamber which is 12\,cm high and has a diameter of 40\,cm. Pipes are connected to 64 hose connections arranged on two rows at the periphery of the chamber. A solenoid valve is mounted on each pipe to individually turn on or off each jet according to a random {spatio-temporal} pattern (see {Sect.~\ref{sec:algo}} below). The other pipe ends are connected to the glass tank from below and are arranged around a 80\,mm-diameter hole [see Fig.~\ref{fig:setup_turbulence}\,(c)]. This central hole is connected to the inlet port of the pump through a PVC (PolyVinyl Chloride) pipe equipped with a flow meter so that the water is in a closed system. A PVC slab is placed 25\,cm above the bottom {of the tank} and is connected to {it} with pipes to redistribute the jets. This way, they are in a $8 \times 8$ array, with a 5\,cm square mesh size and a 16.5\,mm inside diameter [see Fig.\,\ref{fig:setup_turbulence}\,(b)]. Suction holes of 25\,mm in diameter are carved in the PVC slab at the center of each square formed by inlet jets.

By initially filling the tank, the water depth can be adjusted until 75\,cm, which is the distance from the plate with the jets to the top of the tank. In the experiments described in this article, the water depth is set to 66\,cm. Like other RASJA experiments, this device is a closed hydraulic circuit and the amount of water in the tank is constant.
\rev{The main difference} in our experimental setup \rev{is that} we {can} tune the level of turbulence by changing the flow rate of the centrifugal pump. 
{
Our pump 7.5\,kW Grundfos NB 65-125/137 (AFA BAQE) is constituted of an asynchronous motor, whose rotation rate is set by a frequency driver (Inorea Powtran PI8100 11\,kW). The pump can reach a pressure of 2.4\,bars or a flow rate of 38\,L s$^{-1}$ at a pressure of 1.6\,bars. The flow rate is controlled using a feedback loop with a flow meter (Mecon mag-flux A).} 

\subsection{Algorithm to produce random spatiotemporal jet opening}
\label{sec:algo}

Solenoid valves make each jet independent so that they can be turned on or off according to a random spatio-temporal pattern to produce hydrodynamic turbulence. We chose internally piloted solenoid valves Irritrol 700B-.75 Ultra Flow, which display large orifices and thus low-pressure drops (e.g., 0.15\,bar for a 1-L s$^{-1}$ flow rate). A small plastic piece inside the solenoid valve, initially aiming at avoiding a water hammer effect, was removed to decrease the opening and closing times of the valves to about 1\,s. While we use 64 identical solenoid valves, we observe variability in the individual flow rates probably due to serial effects. For example, a standard deviation of $0.042$\,L s$^{-1}$ was measured for a jet mean flow $Q_j=0.13$\,L s$^{-1}$~\cite{JaminPhd2016}. The jets were arranged in a configuration for which the distribution of flow rates is symmetrical and as homogeneous as possible so that we avoid local mean flows.
Solenoid valves are powered with two 400-W power transformers with a 24-VAC and 50-Hz voltage. They are triggered with two Arduino Mega 2560 microcontrollers and solid-state relays (Measurement Computing SSR-4-OAC-05).

{To optimize turbulence intensity and minimize local mean flows, opening and closing times of each solenoid valve are set by an algorithm, for which only a fraction of the 64 jets is simultaneously open. In their paper, Variano and Cowen~\cite{VarianoJFM2008} developed and used the ``sunbathing algorithm'', for which each jet is independently governed by randomly picking up successive opening and closing durations. This algorithm displays low mean flow, high Reynolds number and good spatial homogeneity and has been used in most of subsequent RASJA experiments~\cite{BellaniEF2014,CarterEF2016, JohnsonJFM2017,EstebanJFM2019}. While the average number of jets firing at the same time is chosen through average opening and closing durations, the number of jets firing at any instant is not constant. However, as our experimental setup involves a unique centrifugal pump, the flow rate should be kept constant. Thus, in this paper, we use a pattern very close to the constant-momentum input algorithm mentioned by Variano and Cowen~\cite{VarianoJFM2008} who report performance similar to the ``sunbathing algorithm''. In our algorithm, the number of jets {$N_{on}$} firing at the same time remains constant and opening times are chosen within a continuous uniform distribution {$f_{on}$} of mean value $\mu_{on}$ and half-width $\sigma_{on}$. Initially, $N_{on}$ jets are randomly chosen. Each one is open during a time $d_{on}$ randomly chosen between 0 and $\mu_{on}{\pm\sigma_{on}}$ (uniform distribution). Each time a jet closes, it is replaced by another jet randomly and uniformly picked up from closed ones (no predefined spatial pattern). The opening time $d_{on}$ of this new jet is randomly chosen from the continuous uniform distribution $f_{on}$.}

As a solenoid valve needs around 1\,s to reach a satisfying flow rate, we choose $\mu_{on}=3\,$s and $\sigma_{on}=1\,$s like Variano and Cowen~\cite{VarianoJFM2008}. These authors also found that the highest level of turbulence is reached for $N_{on}=8$ jets open at the same time. Due to the variability in the individual flow rates of the solenoid valves mentioned above, we choose $N_{on}=16$ to keep the global flow rate more constant in time and homogeneous in space.

\subsection{Measurement techniques}
The velocity field under the free surface is measured using Particle Image Velocimetry (PIV). A continuous laser sheet passing through the basin center illuminates a vertical slice of water seeded with 20-$\upmu$m Polyamid Seeding Particles (PSP). A high-speed camera (Phantom V10) {images a two-dimensional domain in the illuminated region with a resolution of $2400 \times 720$ pixels.} {In the following, the measurement domain located in the vertical plane at $y=0$ corresponds to $x \in [-15,15]$\,cm for the horizontal coordinate and to $z \in [-9,0]$ cm for the vertical coordinate. $z=0$ is the position of the top free surface and the point $(x=0,~y=0)$ is the {horizontal} center of the square tank.} {Spatial measurements needing a statistical convergence are} obtained by computing one velocity field per second for $1800$\,s, with a $2$-ms time interval between images of a pair.
Velocity vectors are computed using the PIVLab plugin for Matlab \cite{PIVlab} in a grid of $4 \times 4$-mm$^2$ windows {with a} 2\,mm {mesh size}. We denote $u$ and $w$ the instantaneous local velocities in the horizontal (along $x$ coordinate) and vertical (along $z$ coordinate) directions of the measurement plane, respectively. Their local time-averaged values are ${U}(x,y,z)=\overline{u(x,y,z,t)}$ and ${W}(x,y,z)=\overline{w(x,y,z,t)}$, respectively. The velocity component $v$ (along $y$ coordinate) is perpendicular to the measurement plane and it is thus not measured.
Velocity fluctuations are characterized {by $u'=u-U$ and $w'=w-W$, and their respective standard deviations $\sigma_u$ and $\sigma_w$, with $\sigma_u=\sqrt{\overline{u'^2}}=\sqrt{\overline{u^2}-{U}^2}$}.

The characterization of the velocity field is completed with Laser Doppler Velocimetry (LDV) measurements performed with a Dantec Dynamics FlowExporer 1D apparatus. This local measurement is also performed by seeding water with 20-$\mu$m {PSP}. The velocity temporal signal is interpolated to obtain a constant sampling frequency of ${1000}$\,Hz. All LDV measurements are performed in $x=0$ and for various $z$, while two positions are used for $y$ ($0$ and $-7$\,cm), the uncentered one being able to reach smaller depths for vertical measurements as it is less restricted by the constraint of non-intersection between the laser beams and the free surface.

\begin{figure}
\centering
\includegraphics[width=8.8cm]{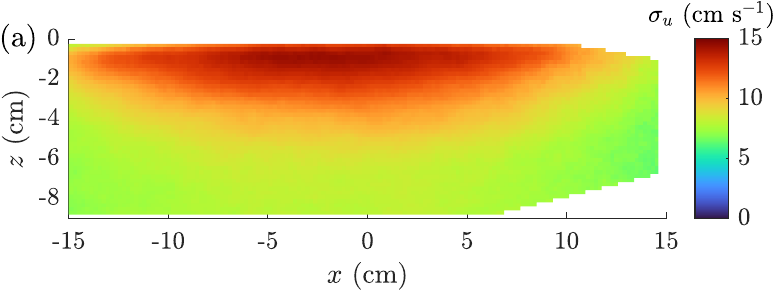}
\includegraphics[width=8.8cm]{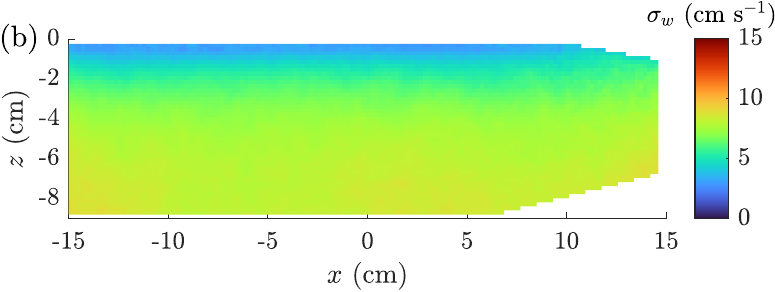}\\
\includegraphics[width=8.8cm]{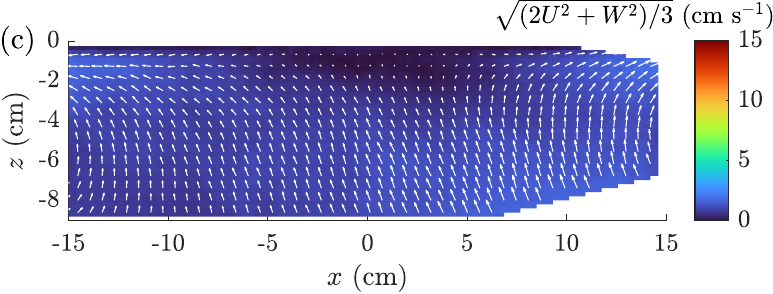}
\includegraphics[width=8.8cm]{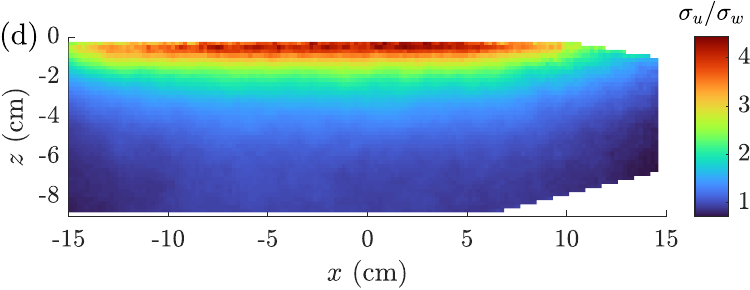}
\caption{Standard deviations of (a) horizontal and (b) vertical velocity fluctuations, $\sigma_u$ and $\sigma_w$ respectively{, (c) mean velocity (absolute value in color and direction as arrows), (d) anisotropy ratio $\sigma_u / \sigma_w$} in the vertical plane at $y=0\,$cm using Particle Image Velocimetry (PIV). White regions are regions not lighted by the laser sheet. Parameters of Experiment A. The free surface is located at $z=0$.}
\label{fig:sigma_u2Ddeform}
\end{figure}

\subsection{General features of generated turbulent flows}
\label{sec:sigmaRHI}
The turbulent flows generated with our device correspond to $16$ jets (over $64$) open at the same time, each remaining open for an average time of $3\,$s, with a mean flow rate per jet $Q_j$ varying between $0.06$ and $0.56$\,L s$^{-1}$ and a water depth of $66\,$cm. In these conditions, we obtain turbulent flows with a low mean flow, a satisfying homogeneity, and a good isotropy except in the vicinity of the jets and of the free surface, similar to the previous RASJA experiments. Colormaps of velocity fluctuations $\sigma_u$ and $\sigma_w$ are displayed in Figs~\ref{fig:sigma_u2Ddeform}{\,(a) and (b)} for the Experiment A, a typical case corresponding to a flow rate per jet $Q_j = 0.375$\,L s$^{-1}$ (characteristics are available in Table~\ref{tab:caract_turb}).
Notice that the top and bottom-right corners are not illuminated so full-width measurements are available for $z\in[-7;-1.2]\,$cm only.

\begin{figure}
\centering
\includegraphics[width=8.8cm]{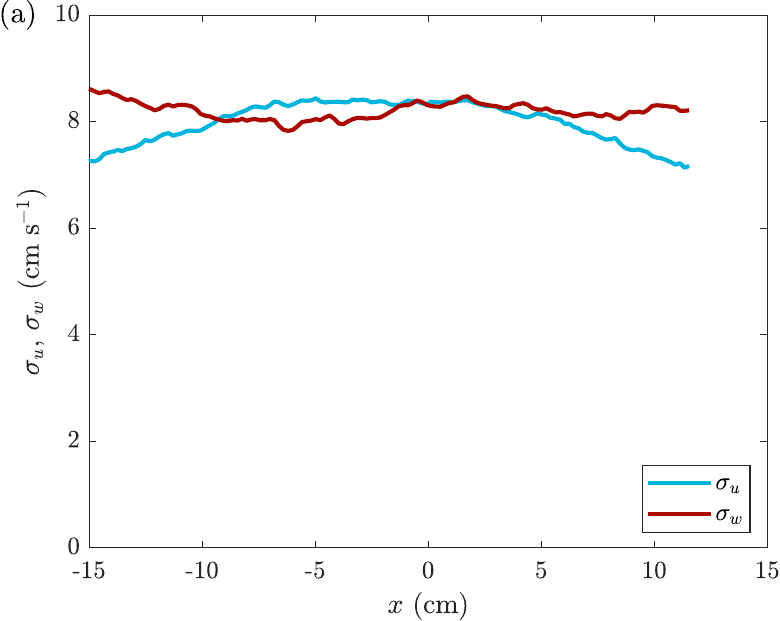}
\hfill
\includegraphics[width=8.8cm]{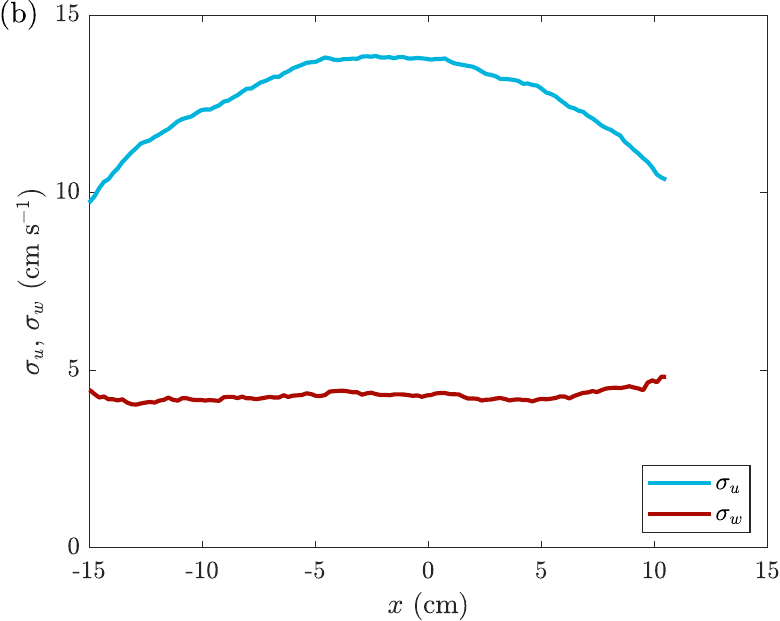}
\caption{Horizontal profiles of {standard deviations for horizontal and vertical} velocity fluctuations {$\sigma_u$ and $\sigma_w$}, averaged vertically (a) in the nearly-homogeneous region ($-7.4 \leq z \leq -6\,$cm) and (b) close to the free surface ($-1.7 \leq z \leq -0.2\,$cm). Parameters of Experiment A.}

\label{fig:profil_spatial_horizontal}
\end{figure}

The region of interest is sufficiently far from the bottom plate, to not observe the signature {of individual jets in the velocity field}. {The mean flow is represented on Fig.\,\ref{fig:sigma_u2Ddeform}\,(c), with its absolute value $\sqrt{(2\,U^2+W^2)/3}$ in color and its direction with arrows. We see that it is upward but with low values compared to fluctuations.} We observe that horizontal velocity fluctuations are maximal near the free surface and the horizontal center of the tank while vertical velocity fluctuations decrease when approaching the free surface. {This is highlighted in Fig.\,\ref{fig:sigma_u2Ddeform}\,(d) displaying the anisotropy ratio, which is close to 1 far from the free surface and strongly increases near the free surface.} {Figure~\ref{fig:profil_spatial_horizontal}\,(a) shows the horizontal velocity profile, computed from Figs~\ref{fig:sigma_u2Ddeform}{\,(a) and (b)} {averaged over $z\in[-7.4;-6]\,$cm}. $\sigma_u$ and $\sigma_w$ display good {isotropy and} homogeneity, in particular for $-5<x<5\,$cm. When coming closer to the free surface {(see Fig.\,\ref{fig:profil_spatial_horizontal}\,(b), same as (a) but averaged over $z\in[-1.7;-0.2]\,$cm)}, the homogeneity stands for $\sigma_w$ while lateral boundaries affect $\sigma_u$. This latter phenomenon is because eddies are horizontally elongated}, as will be studied {and detailed} in Sect.~\ref{sec:nearFS}.

\begin{figure}
\centering
\includegraphics[width=11cm]{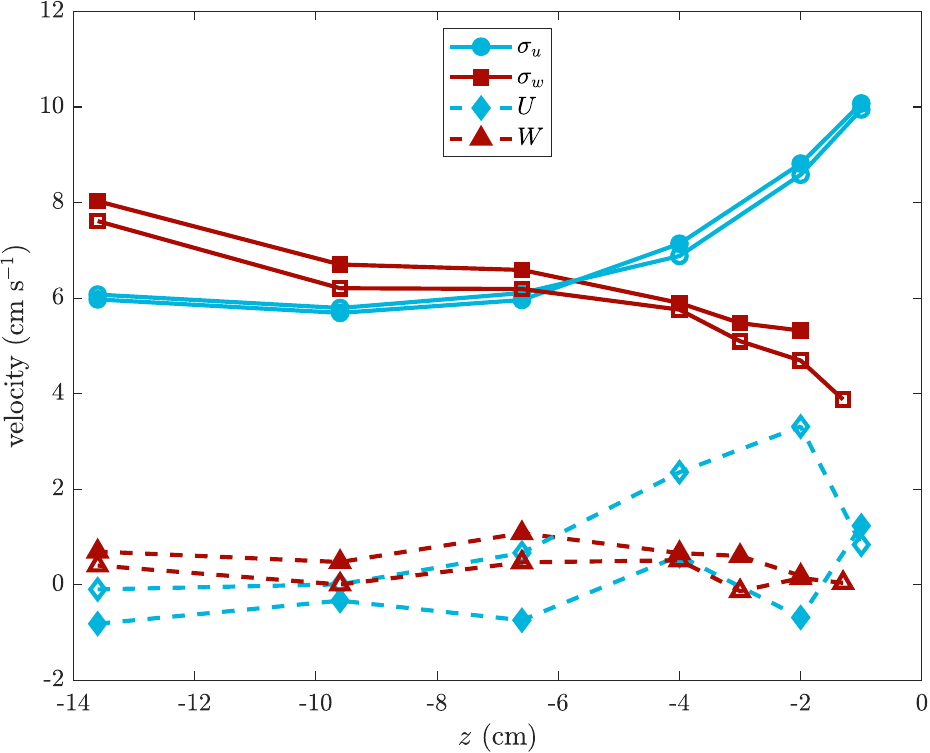}
\caption{LDV measurements of {standard deviations $\sigma_u$ and $\sigma_w$ and mean flows $U$ and $W$ for horizontal and vertical velocities} as functions of $z$. Each point is an independent measurement of 1200-s duration. $Q_j=0.25$\,L s$^{-1}$, $x=0$ and $y=0$ (filled symbols) or $y=-7\,$cm (empty symbols).}
\label{fig:profil_spatial_vertical_vitesses_bis}
\end{figure}

Besides, Fig.\,\ref{fig:profil_spatial_vertical_vitesses_bis} displays the vertical profile of velocity fluctuations using LDV individual measurements, including measurements below the PIV domain (vertical PIV profile will be studied in Sect.~\ref{sec:nearFS}). While turbulent intensity decays when moving away from the forcing region, a plateau is formed for depth between $-10$ and $-6$\,cm, similar to Variano and Cowen~\cite{VarianoJFM2008} observations. The plateau is also consistent with the homogeneous regions visible on the lower parts of Figs~\ref{fig:sigma_u2Ddeform}\,(a) and (b).

\begin{figure}
\centering
\includegraphics[width=11cm]{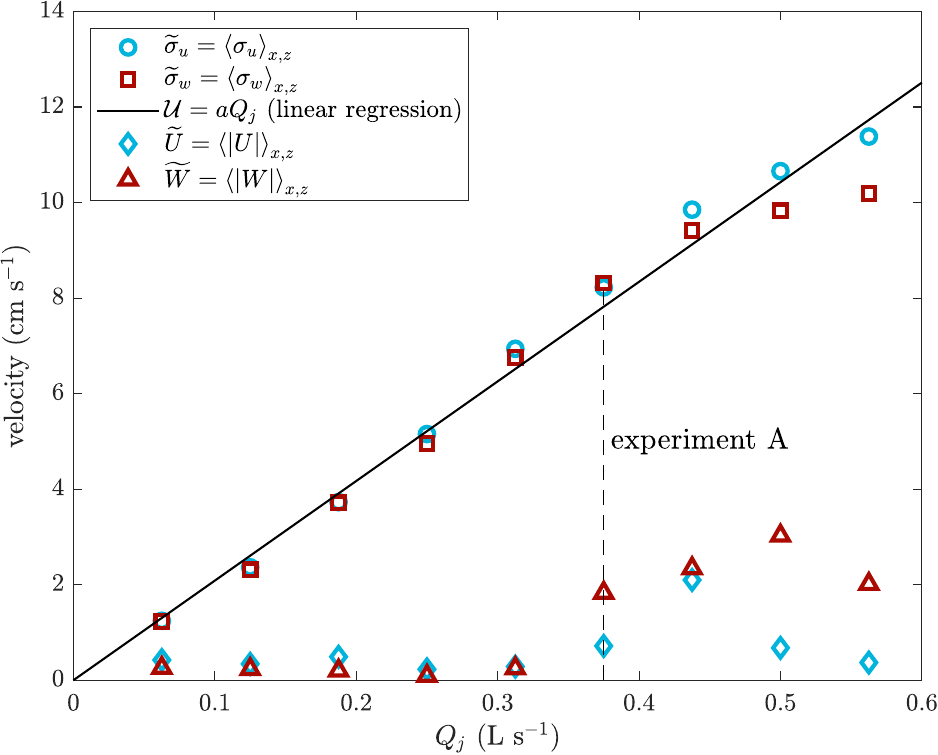}

\caption{Standard deviations ($\widetilde{\sigma}_u$ and $\widetilde{\sigma}_w$) and mean values {($\widetilde{U}$ and $\widetilde{W}$)} of horizontal and vertical velocities, averaged over the homogeneous region defined by $-5<x<+5$\,cm and {$-9<z<-6$\,cm} as a function of the flow rate per jet $Q_j$. Black line corresponds to the linear regression of equation $\mathcal{U}=aQ_j$ with {$a=209$\,m$^{-2}$ (95\,\% CI: [200; 217])}, where $\mathcal{U}=\sqrt{(2\widetilde{\sigma}_u^2+\widetilde{\sigma}_w^2)/3}$.}

\label{fig:sigmaRHI}
\end{figure}

The region defined by $-5<x<+5$\,cm and $-9<z<-6$\,cm displays a very good homogeneity on these figures. Indeed, the standard deviation of $\sigma_u$ and $\sigma_w$ over this region is equal to 2.4\,\% and 2.3\,\% of $\sigma_u$ and $\sigma_w$ respectively when averaging over all measurements (and below 3.6\,\% for each measurement). So the turbulence intensity will now be quantified by averaging $\sigma_u$ and $\sigma_w$ over this region. {In the rest of the paper, values averaged over this region will be noted with a tilde.} The averaged values $\widetilde{\sigma}_u$ and $\widetilde{\sigma}_w$ are plotted as functions of the flow rate per jet $Q_j$ in Fig.\,\ref{fig:sigmaRHI} {and reported in Table~\ref{tab:params}}.

\begin{table}
   \centering
   \begin{ruledtabular}
\begin{tabular}{|l |c |c| c|c |c| c|c |c| c|}
Flow rate per jet $Q_j$ (L s$^{-1}$) & 0.0625  &  0.125 &   0.1875  &  0.25  &  0.3125  &  \textbf{0.375} &   0.4375   & 0.5  &  0.5625 \\
\hline
Horizontal velocity fluctuations $\widetilde{\sigma}_u$ \,(cm s$^{-1}$) &  1.24  &  2.36  &  3.73  &  5.15  &  6.94  & \textbf{8.22}  &  9.85  & 10.66  & 11.38 \\
\hline
Vertical velocity fluctuations $\widetilde{\sigma}_w$ \,(cm s$^{-1}$)&  1.23  &  2.31  &  3.71  &  4.95  &  6.76   & \textbf{8.31}  &  9.42  &  9.83 &  10.19 \\
\hline
{Velocity fluctuations $\mathcal{U}=\sqrt{(2\,\widetilde{\sigma}_u^2+\widetilde{\sigma}_w^2)/3}$} \,(cm s$^{-1}$) &  1.24  &  2.35  &  3.72  &  5.09  &  6.88  &  \textbf{8.25}  &  9.71 &  10.39 &  11.00 \\

\end{tabular}
\end{ruledtabular}
\caption{Parameters and velocity fluctuations of the experiments presented in the paper. Spatial averaging is performed far from the free surface in the homogeneous region defined by $-5<x<+5$\,cm and {$-9<z<-6$\,cm}. Experiment A is in bold.}
\label{tab:params}
\end{table}

We observe that, beyond homogeneity, this region displays a good isotropy, i.e., $\widetilde{\sigma}_u\approx\widetilde{\sigma}_w$, for all the flow rates. The local anisotropy ({$\widetilde{\sigma}_w/\widetilde{\sigma}_u\in [0.90 ; 1.01]$} over all the experiments) is {lower than the value measured} by Variano and Cowen~\cite{VarianoJFM2008} (value of 1.27) or reported by the same authors from others studies (values between 1.1 and 1.4 for grid-stirred tanks). {We can see in Fig.\,\ref{fig:profil_spatial_vertical_vitesses_bis} that the anisotropy {below} the region defined by $-9<z<-6$\,cm is {more important, $\sigma_w$ being larger}. The good local isotropy {mentioned above} is thus obtained {in the intermediary region only} thanks the presence of the free surface which {favors $\sigma_u$ over $\sigma_w$}, as will be studied in Sect.~\ref{sec:nearFS}.}
Besides, $\widetilde{\sigma}_u$ and $\widetilde{\sigma}_w$ increase linearly in Fig.~\ref{fig:sigmaRHI} (except for the largest flow rates) with $Q_j$, following $\widetilde{\sigma}_u\approx\widetilde{\sigma}_w=aQ_j$, where {$a=209$\,m$^{-2}$}. {By symmetry, we assume that the statistical quantities associated to $v$ are identical to those measured for $u$.} Thereafter, the turbulence intensity will be characterized by the 3D mean value $\mathcal{U}=\sqrt{(2\,\widetilde{\sigma}_u^2+\widetilde{\sigma}_w^2)/3}$ \cite{VarianoJFM2008}. {$\mathcal{U}$ is directly related to the turbulent kinetic energy which is equal to $\mathcal{U}^2/2$}. We also observe in Fig.\,\ref{fig:sigmaRHI} that mean flow velocities $\widetilde{U}=\left\langle \left|U\right|\right\rangle_{x,z}$ and $\widetilde{W}=\left\langle \left|W\right|\right\rangle_{x,z}$, averaged over the homogeneous region, keep quite low values, remaining at least 3 times smaller than standard deviation values. \revDEL{If we consider that the 1-Hz acquisition rate ensures independence of individual velocity values, confidence intervals (CI) can be evaluated. Using the bootstrap method, we find that the widths of 95\,\% CI for standard deviations and means are equal to 7\,\% and 9\,\% of $\mathcal{U}$, respectively. Inspired by the RASJA experiment from \cite{VarianoJFM2008}, our setup thus adds the ability to control the turbulence intensity by replacing the 64 fixed-flow pumps with a variable-flow pump and 64 solenoid valves.}

\section{Homogeneous and isotropic turbulence far from the free surface}
\label{Homogeneousturbulence}

In this section, we characterize the turbulent flow far from the free surface {(}depth $-9<z<-6\,$cm{)}, i.e., in the part of the homogeneous region belonging to the PIV measurement domain. In this region, the effect of the free surface is negligible and the turbulence is homogeneous and isotropic, as shown in the previous section. The general features of the generated turbulent flows are summarized in Table~\ref{tab:caract_turb} and compared to the RASJA experiment of Variano and Cowen~\cite{VarianoJFM2008}. The definitions of characteristic numbers and the methods of calculation are then provided in this section.

\begin{table}
   \centering
   \begin{ruledtabular}
\begin{tabular}{|l ||c |c|| c|}
& \, Range of study \, & \, Experiment A \, & \, Variano and Cowen \cite{VarianoJFM2008}\,\\
  \hline
Flow rate per jet $Q_j$ (L s$^{-1}$) & $[0.06 ; 0.56]$  & 0.375& 0.38 \\
Horizontal velocity fluctuations $\widetilde{\sigma}_u$ \,(cm s$^{-1}$) & $[1.24 ; 11.4]$ & 8.22  &  3.91 \\
Vertical velocity fluctuations $\widetilde{\sigma}_w$ \,(cm s$^{-1}$)&  $[1.23 ; 10.2]$  & 8.31  & 4.98 \\
{Velocity fluctuations $\mathcal{U}=\sqrt{(2\,\widetilde{\sigma}_u^2+\widetilde{\sigma}_w^2)/3}$} \,(cm s$^{-1}$) & $[1.24 ; 11.0]$ & 8.25 & 4.30 \\
Anisotropy ratio $\widetilde{\sigma}_w/\widetilde{\sigma}_u$ & $[0.90 ; 1.01]$ & 1.01 & 1.27 \\
Horizontal fluctuation rate $\widetilde{U}/\widetilde{\sigma}_u$ & [0.032; 0.34] &  0.087 & 0.07\\
Vertical fluctuation rate $\widetilde{W}/\widetilde{\sigma}_w $ & [0.018; 0.31] & 0.22  & 0.02 \\
Longitudinal horizontal integral length scale $\widetilde{L}_{uu}$\,(cm) & {[5.5; 7.8] (mean: $6.6$)} & {$7.0$} & 7.57\,\\
Transverse vertical integral length scale $\widetilde{L}_{ww}$\,(cm) & {[3.1; 4.1] (mean: $3.7$)} & {$3.9$} & \, 6.36\,\\
Integral length scale ratio $\widetilde{L}_{uu}/\widetilde{L}_{ww}$\,(cm) & {[1.6; 2.0] (mean: $1.8$)} & {$1.8$} & \, 1.19\,\\
Dissipation rate $\epsilon$ \,(cm$^2$ s$^{-3}$) & $[0.16 ; 71]\,$ & 33 & 5.20 \\
Kolmogorov length scale $\eta_K$\,(mm) & $[0.11 ; 0.50]\,$ & 0.13 & 0.21\\
Kolmogorov time scale $\tau_\eta$\,(ms) & $[11.9 ; 250]\,$ & 17.5 & 44\\
Kolmogorov velocity scale $v_\eta$\,(mm s$^{-1}$) & $[2.0 ; 9.2]\,$ & 7.6 & 4.8\\
Turbulent Reynolds number  $\mathrm{Re}_T$ & $[991 ; 8801]$ & 6603 & 3250 \\
Taylor Reynolds number  $\mathrm{Re}_\lambda$ & $[148 ; 558]$ &461 & 314
\end{tabular}
\caption{Features of the turbulence in the homogeneous region generated with our RASJA device. First column: range of values measured in this study. Second column: values in the typical experiment A. Third column: values obtained with RASJA experiment by Variano and Cowen \cite{VarianoJFM2008}, presented for comparison.}
\label{tab:caract_turb}
\end{ruledtabular}
\end{table}

\subsection{Integral length scales}
\label{sec:echelles}

\begin{figure}
\centering

\includegraphics[width=8.8cm]{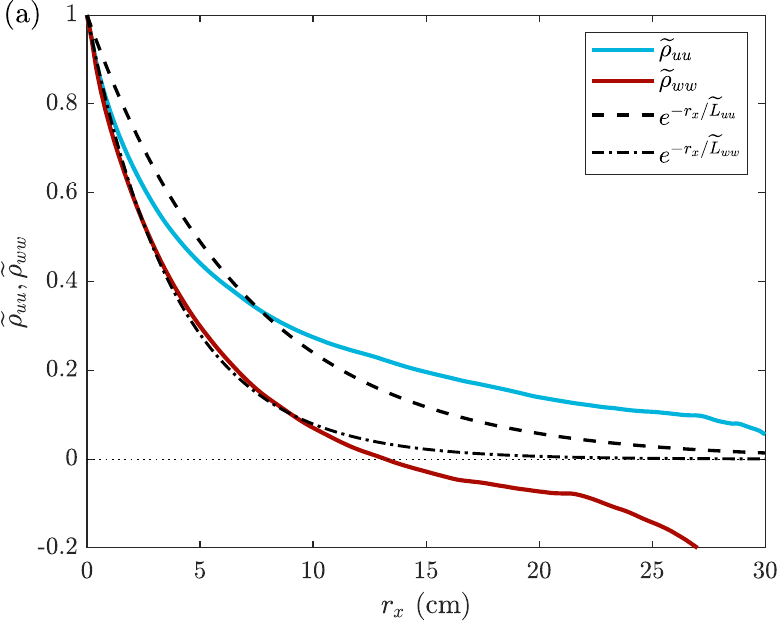}
\hfill
\includegraphics[width=8.8cm]{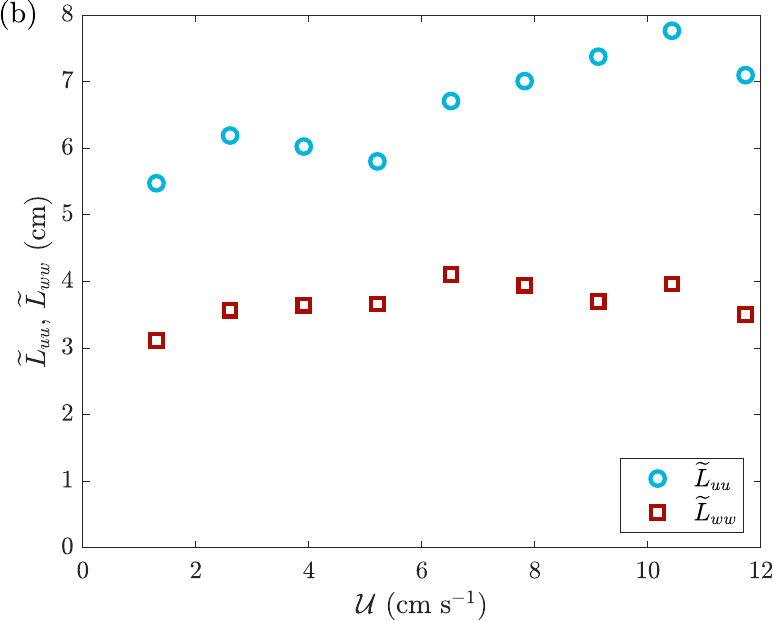}

\caption{(a) Autocorrelation functions $\widetilde{\rho}_{uu}$ and $\widetilde{\rho}_{ww}$ (light blue and dark red solid lines, respectively) {of horizontal and vertical velocities $u$ and $w$ along $x$} {and corresponding exponential fittings (black dashed and dash-dottted lines, respectivety)} as functions of spatial increment $r_x$. Parameters of Experiment A; $y=0$ and averaged over $z\in[-7;-6]\,$cm (full-width illuminated region). (b) Longitudinal $\widetilde{L}_{uu}$ and transverse $\widetilde{L}_{ww}$ integral scales {of horizontal and vertical velocities respectively} versus velocity fluctuations $\mathcal{U}$. Integral scales are obtained by integration over $r_x$ of curves fitting $\widetilde{\rho}_{uu}$ and $\widetilde{\rho}_{ww}$ as shown in (a).}

\label{fig:LvsQj7}
\end{figure}

Integral length scales characterize the size of the largest eddies and may be computed by the integral of the spatial autocorrelation function for velocities~\cite{Pope}. A relevant integral length scale here is the one involving $u$ along the \rev{horizontal} axis $O_x$, for $y=0$. The corresponding {normalized} autocorrelation function and averaged over time is~\cite{Pope}:
{\begin{equation}
\rho_{uu}(r_x,y,z)=\overline{\left[\frac{\left\langle u'(x,y,z,t) u'(x+r_x,y,z,t) \right\rangle_x}{\left\langle u'(x,y,z,t) \right\rangle_x^2}\right]}
\label{eq:aurx}
\end{equation}}

\noindent where spatial increment $r_x$ is aligned with $Ox$. The longitudinal integral length scale for the horizontal velocity is {$L_{uu}(y,z)=\int_0^\infty \rho_{uu}(r_x,y,z)dr_x$}. In the region far from the free surface, we define the normalized autocorrelation function {$\widetilde{\rho}_{uu}(r_x)=\left\langle\rho_{uu}(r_x,y,z)\right\rangle_z$, taken in $y=0$ and} averaged over vertical space ($-9 <z <-6$\,cm), and its associated integral length scale $\widetilde{L}_{uu}=\int_0^\infty \widetilde{\rho}_{uu}(r_x)dr_x$. The transverse autocorrelation function for the vertical velocity $\rho_{ww}(r_x,y,z)$ and its corresponding quantities are defined in the same way by replacing $u$ by $w$ in the above equations. $\widetilde{\rho}_{uu}$ and $\widetilde{\rho}_{ww}$ are represented in Fig.\,\ref{fig:LvsQj7}\,(a).

{We observe over all our experiments that $\widetilde{\rho}_{ww}$ can be very well approximated by an exponential fit $e^{-r_x/\widetilde{L}_{ww}}$ for $r_x<15\,$cm [black dash-dotted line in Fig.\,\ref{fig:LvsQj7}\,(a)]. In contrast, the fit of  $\widetilde{\rho}_{uu}$ by an exponential fit  $e^{-r_x/\widetilde{L}_{uu}}$ is average. For the experiment A, we find with the exponential fits, $\widetilde{L}_{ww}=4.2$ cm and $\widetilde{L}_{uu}=7.0$ cm. We note that the separation length where the autocorrelation is equal to $1/e$ gives very similar values, but a little less converged {(not shown here)}. We also observe that $\widetilde{L}_{uu}$ increases with the turbulence strength, while $\widetilde{L}_{ww}$ do not change significantly, even though it is compatible with a slight increase [Fig.\,\ref{fig:LvsQj7}\,(b)]. Here, we are interested in finding an order of magnitude of the longitudinal integral length. The value for $\widetilde{L}_{uu}$ is not well defined as the range in $x$ is limited and the exponential fit is average, so we assume from now that the integral scale ratio $\widetilde{L}_{uu}/\widetilde{L}_{ww}$  is $2$ as expected in homogeneous and isotropic turbulence~\cite{Pope}. Thereafter, we use the more reliable value of $\widetilde{L}_{ww}$, which is $3.7\,$cm $\approx 4\,$cm on average, to compute the integral length scale of the flow $L= \widetilde{L}_{uu}\approx8\,$cm, which characterizes the size of the largest eddies.}

\begin{figure}
\centering
\includegraphics[width=8.8cm]{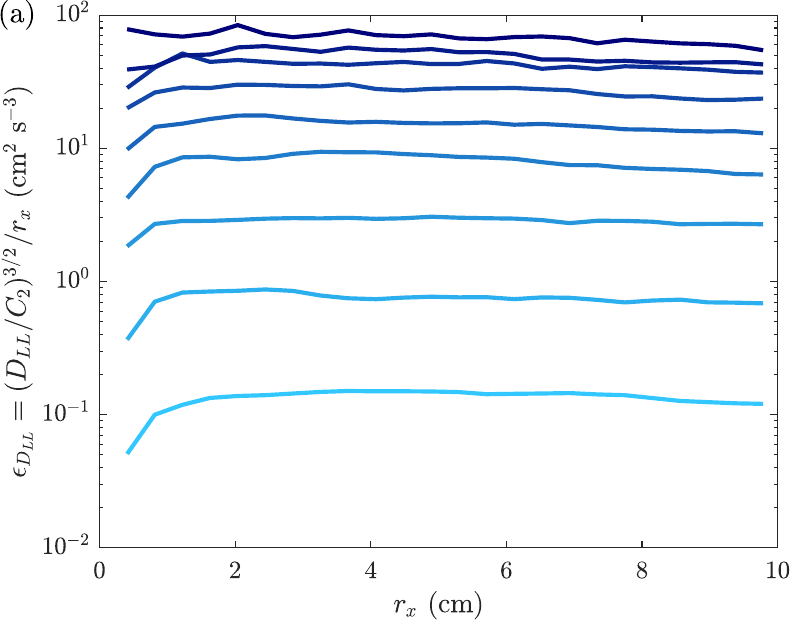}
\hfill
\includegraphics[width=8.8cm]{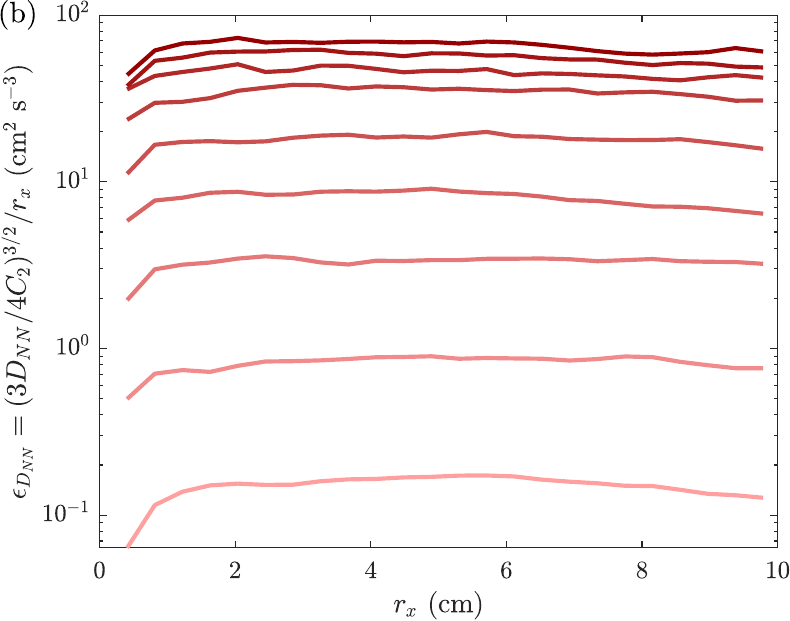}

\caption{Dissipation rates $\epsilon_{D_{LL}}$ and $\epsilon_{D_{NN}}$ computed from longitudinal and transverse second-order structure functions (a) $D_{LL}$ for horizontal velocity $u$ and (b) $D_{NN}$ for vertical velocity $w$ as functions of the spatial increment $r_x$. {Darkest lines \ correspond to highest} levels of turbulence, $\mathcal{U}$ {See Table~\ref{tab:params}}. $y=0$, $z=-6.4\,$cm.}

\label{fig:epsilonDLL}
\end{figure}

\begin{figure}[h!]
\centering
\includegraphics[width=11cm]{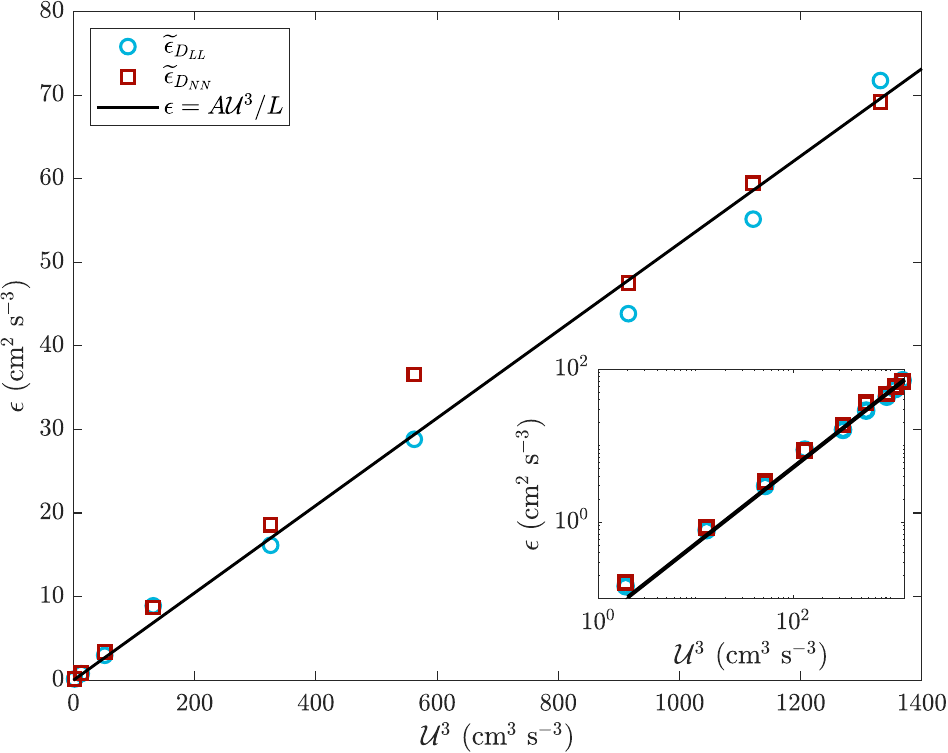}

\caption{Estimations of dissipation rates $\widetilde{\epsilon}_{D_{LL}}$ and $\widetilde{\epsilon}_{D_{TT}}$ as functions of turbulent velocity fluctuations $\mathcal{U}^3$, obtained by averaging over spatial increment $r_x$ dissipation rates plotted in Fig.\,\ref{fig:epsilonDLL}. Solid line corresponds to the scaling law $\epsilon=A\mathcal{U}^3/L$, with {$A=0.42$ (95\,\% CI: [{0.40}; 0.43])} and $L=8$ cm. Inset: same as the main plot but in logarithmic scale.}

\label{fig:epsilonsvsU}
\end{figure}

\subsection{Dissipation rate}
\label{sec:epsilon}

The dissipation rate $\epsilon$ is associated with the energy injected at large scales, which goes through inertial scales and is finally dissipated at small scales. It is defined by $\epsilon = 2\,\nu \left\langle S_{ij}S_{ij}\right\rangle$, where {$S_{ij}=\left(\partial u'_i/\partial x_j+\partial u'_j/\partial x_i\right)/2$} is the velocity gradient and $\nu=1\ 10^{-6}$m$^2$~s$^{-1}$ the kinematic viscosity of water. Our PIV resolution is not large enough to compute $\epsilon$ from this definition. {But it can be computed through time-averaged second-order structure functions along $x$~\cite{Pope}:}

{\begin{eqnarray}
D_{LL}(r_x,y,z) & = &\overline{\left\langle\left[u'(x,y,z,t)-u'(x+r_x,y,z,t)\right]^2\right\rangle}_{x} \\
D_{NN}(r_x,y,z) &= & \overline{\left\langle\left[w'(x,y,z,t)-w'(x+r_x,y,z,t)\right]^2\right\rangle}_{x}
\end{eqnarray}}
\noindent where $r_x$ is the spatial increment along $Ox$ axis. In the inertial range, $D_{LL}=C_2(\epsilon \, r_x)^{2/3}$ and $D_{NN}=\frac{4}{3} C_2(\epsilon \, r_x)^{2/3}$, with $C_2=2.0$~\cite{Pope}. We computed the associated dissipation rate $\epsilon_{D_{LL}}$ and $\epsilon_{D_{NN}}$ for various $r_x$. 
The obtained values display a good homogeneity over the whole inertial range as shown in Fig.\,\ref{fig:epsilonDLL} {for $y=0$ and $z=-6.4\,$cm}. By averaging them over the interval $2<r_x<6\,$cm, we obtain the estimations for dissipation rate $\widetilde{\epsilon}_{D_{LL}}$ and $\widetilde{\epsilon}_{D_{NN}}$, which have similar values and are plotted in Fig.\,\ref{fig:epsilonsvsU} as functions of $\mathcal{U}^3$. The linear shape of the curve is compatible with the scaling law $\epsilon=A\mathcal{U}^3/L$, with $A=0.42$.  The value of $A$ is of the same order of magnitude of the one found by Variano and Cowen \cite{VarianoJFM2008} ($A\approx 0.52$). In the following, the dissipation rate is $\epsilon=(\widetilde{\epsilon}_{D_{LL}}+\widetilde{\epsilon}_{D_{NN}})/2\in [0.16 ; 71]\,$cm$^2$ s$^{-3}$ .

These values of $\epsilon$ give us access to Kolmogorov length, time and velocity scales, {$\eta_K=(\nu^3/\epsilon)^{1/4}\in[0.11 ; 0.50]\,$mm, $\tau_\eta=(\nu/\epsilon)^{1/2}\in[11.9 ; 250]\,$ms and $v_\eta=(\nu\epsilon)^{1/4}\in[2.0;9.2]\,$mm.s$^{-1}$}, which are characteristic scales of the smallest turbulent motions~\cite{Pope}.

\subsection{Velocity power spectra}
\label{sec:spectres_caract}

\begin{figure}
\centering
\includegraphics[width=11cm]{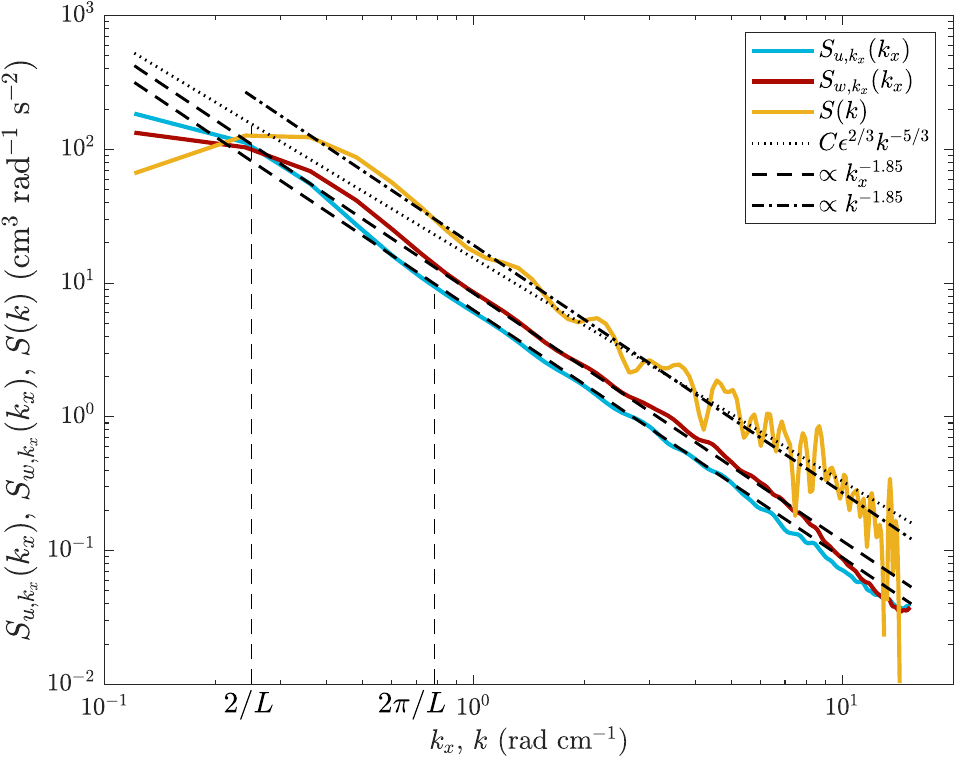}

\caption{Longitudinal (light blue),  transverse (dark red){, and 3D (yellow) spatial power spectra of turbulent velocity fluctuations {$S_{u,k_x}(k_x)$, $S_{w,k_x}(k_x)$}, and $S(k)$ as functions of horizontal wavenumber $k_x$ and wavenumber $k$} far from the free surface and averaged over time. Dotted line represents the theoretical longitudinal spectra $C'_\parallel\epsilon^{2/3}k_x^{-5/3}$. {Dash-dotted line and dashed lines correspond to the scaling law in $k^{-1.85}$. Dashed lines are obtained by multiplying the prefactor of the dash-dotted line by $C'_\parallel/C=18/55$ and $C'_\perp/C=24/55$.} Parameters of Experiment A; $y=0$ and $z=-6.4\,$cm.}

\label{fig:spectre_spatial_u_w_-6_4cm}
\end{figure}

\begin{figure}
\centering
\includegraphics[width=11cm]{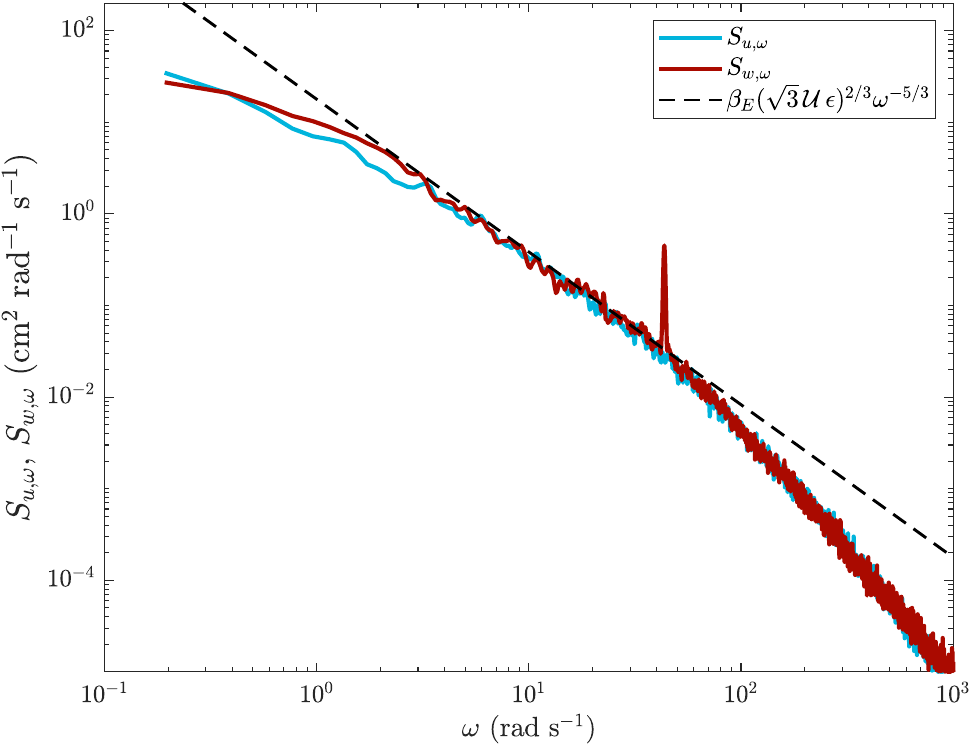}

\caption{Temporal power spectra $S_{u,\omega}$ and $S_{w,\omega}$ of the horizontal (light blue) and vertical (dark red) velocity as functions of the angular frequency $\omega$ for $x=0$, $y= -7$\,cm and $z=-6.6\,$cm with two distinct LDV measurements for {horizontal and vertical velocities $u$ and $w$}. {Dashed line represents the theoretical law predicted by Tennekes \cite{TennekesJFM1975} choosing $\beta_E=0.7$.} Duration: 1200\,s. $Q_j=0.25\,$L s$^{-1}$. $\mathcal{U}=6.1\,$cm s$^{-1}$.}
\label{fig:spectre_temporel_u_w}
\end{figure}

Homogeneous and isotropic turbulence is a scale-invariant phenomenon, which is thus well described in the spatial Fourier space. The power-law behavior of the {three-dimensional} velocity power spectrum $S(k)= C\,\epsilon^{2/3}\,k^{-5/3}$ in the inertial range (predicted by Kolmogorov in 1941) constitutes one of the most robust results in homogeneous and isotropic turbulence~\cite{TennekesJFM1975,FrischBook,Pope,Bailly}. $k$ is the spatial wavenumber and $C\approx 1.5$ is the experimentally measured Kolmogorov constant~\cite{Pope,SreenivasanPoF1995}. Normalized spatial power spectra of turbulent velocity fluctuations are computed along the horizontal coordinate $x$ for {$u'$ and $w'$ on instantaneous velocity fluctuations} multiplied by Blackman–Harris window functions. The squared modulus of the fast Fourier transform is normalized and averaged over time, yielding longitudinal and transverse spectra, $S_{u,k_x}$ and $S_{w,k_x}$.
{From these spectra, we derive the  three-dimensional velocity power spectrum $S(k)$ with the assumption of isotropy using the relation~\cite{Pope}:}
\begin{equation}
{S(k)=-k\,\dfrac{\mathrm{d}}{\mathrm{d} k} \left[ \frac{1}{2}\, S_{u,k_x} (k) + S_{w,k_x} (k) \right]  \, .}
\end{equation}

 {We display $S_{u,k_x}$, $S_{w,k_x}$, and $S$ in Fig.~\ref{fig:spectre_spatial_u_w_-6_4cm} for $z=-6.4\,$cm. Interestingly, we find that the maximum of $S(k)$ at large scales is located approximately at $k=2/L$, \textit{i.e.} the longitudinal integral scale corresponds well to the most energized scales, from which the direct cascade starts. The black dotted line in Fig.~\ref{fig:spectre_spatial_u_w_-6_4cm} denotes the theoretical Kolmogorov 41 power-law spectrum $C\epsilon^{2/3}k^{-5/3}$, using $\epsilon$ from Sect.~\ref{sec:epsilon} and $C=1.5$, and shows a rough agreement with the 3D spatial spectrum  (yellow line). Indeed, we find that all our experimental spatial spectra show a slightly steeper scaling law ($\propto k^{-1.85}$, represented by a dash-dotted line and dashed lines) in their sub inertial ranges, as observed in the experiment of Variano and Cowen~\cite{VarianoJFM2008} for 1D spectra. Due to isotropy, a $\epsilon^{2/3}\,k_x^{-5/3}$ power law is also expected in the sub inertial range for $S_{u,k_x}$ and $S_{w,k_x}$ but with different dimensionless prefactors, $C'_\parallel=18/55\,C$ and $C'_\perp=24/55\,C$ for the longitudinal and transverse spectra, respectively~\cite{Pope}. The relative position of observed longitudinal and transverse spectra is consistent with the hypothesis of homogeneous and isotropic turbulence~\cite{Pope}: $S_{u,k_x}<S_{w,k_x}$ in the inertial range while $S_{u,k_x}>S_{w,k_x}$, for the smallest wavenumbers.  Dashed lines in Fig.\,\ref{fig:spectre_spatial_u_w_-6_4cm} are obtained by multiplying the prefactor of the dash-dotted line by $C'_\parallel/C=18/55$ and $C'_\perp/C=24/55$. Their good agreement with experimental 1D spectra shows that these ratio predicted by Kolmogorov 41 are verified.}

The temporal {velocity power} spectra are displayed in Fig.\,\ref{fig:spectre_temporel_u_w}, obtained from LDV measurements for {$z=-6.6\,$cm and for experimental parameters close to those of Experiment A}. We observe a $\omega^{-5/3}$ {power-law spectrum} in the inertial range for $u$ and $w$ {for $\omega \in [2,50]\,$rad~s$^{-1}$}. This {power-law spectrum is} theoretically predicted {using the random sweeping model of Tennekes}~\cite{TennekesJFM1975} and is derived from the Kolmogorov 41 spectrum in space ($\sim k_x^{-5/3}$). The physical origin is similar to Taylor's frozen turbulence hypothesis, valid in the case of turbulence with a mean flow. {Here, in {the} absence of a global mean flow, the scaling law is explained by the advection of the small eddies (inertial range) and of their spatial structure by the largest eddies.} Temporal fluctuations in the corresponding frequencies are thus similar to spatial fluctuations. Tennekes thus predicts that the spectral density is {$S_{u,\omega}=S_{w,\omega}=\beta_E(\sqrt{3}\,\mathcal{U}\,\epsilon)^{2/3}\omega^{-5/3}$, with $\beta_E\sim 1$}. In Fig.\,\ref{fig:spectre_temporel_u_w}, by using the relation $\epsilon=A\mathcal{U}^3/L$ determined from Fig.\,\ref{fig:epsilonsvsU} and by choosing $\beta_E=0.7$, we find that this theoretical equation fits very well to our data.
We note that the vertical velocity spectrum $S_{u,\omega}$ displays a peak around {$\omega=40-50\,$rad~s$^{-1}$}\revDEL{, probably due to mechanical oscillations of the setup, this frequency being quite independent of forcing. However, our setup is not excited at particular frequencies other than this one, contrary to oscillating-grid turbulence experiments. Thus, the inertial range of our temporal spectra in Fig.\,\ref{fig:spectre_temporel_u_w} is not polluted by numerous peaks due to the oscillation frequency of a grid and its harmonics}. \rev{At the difference of oscillating-grid turbulence experiments displaying numerous peaks at the excitation frequency and its harmonics, the frequency corresponding to the peak we observe is quite independent of forcing and is thus probably due to mechanical oscillations of the setup}~\cite{BrumleyJFM1987,chiapponi2012experimental}.

\subsection{Reynolds numbers}
\label{sec:Reynolds}

\begin{table}
   \centering
   \begin{ruledtabular}
\begin{tabular}{|l ||c |c|c|c|c|c|c||}
& \,  Facility dimensions \, & \, $\sigma_u$, $\mathcal{U}$ \, & \, $\dfrac{\sigma_w}{\sigma_u}$ & \,  $L_{uu}$, $L$ \, &  $Re_T$ & $Re_\lambda$ & \, $\epsilon$ \\
& \, L $\times$ W $\times$ H  (cm$^3$) \, & \, (cm s$^{-1}$) \, &    \,  & (cm) &   &    & (cm$^2$ s$^{-3}$) \\
  \hline
Brumley \& Jirka 1987 \cite{BrumleyJFM1987}$^a$ & $50  \times 50 \times 40$ & 0.7 & 1  &  2.5 & 1000 &  112 & $0.1 $ \\
\, \textit{Vertical oscillating grid, free-surface} & grid depth  25 cm & & &  & & & \\
\hline
Herlina \& Jirka 2008 \cite{HerlinaJFM2008}$^a$ & $50  \times 50 \times 65$ & 1.4  & \textit{NA}   &  2.8 & 780 &  \textit{NA} & $\sim 0.5$\\
\, \textit{Vertical oscillating grid, free-surface} & grid depth  28 cm & & &  & & & \\
  \hline
Variano \& Cowen 2008 \cite{VarianoJFM2008}\, & $80 \times 80 \times 100$ &  4.30 & 1.27  & 7.57 & 3250 & 314 & 5.20  \\
  \,  \textit{RASJA, free-surface} & & &  & & & &  \\  
   \hline
 Carter et al. 2016 \cite{CarterEF2016,CarterJFM2017} \, &  $110 \times 200 \times 200$ & 76 & 1.72     & 14 & $10^5$ & 496& $1.2 \times 10^4$     \\
  \, \textit{RASJA, air} & & & &  & & & \\
  \hline
    Johnson \& Cowen 2018  \cite{JohnsonJFM2017} \,   & $80 \times 80 \times 100$ & 5.19 & 1.35   & 9.71 & 5000  & 378 & 12.19   \\
\,\textit{RASJA, water, top} & & \textit{5.97}  & \textit{1.3}  & \textit{5.19} & \textit{2700} & \textit{262} & \textit{36}  \\
  \hline
  Esteban et al. 2019 \cite{EstebanJFM2019} \,   & $85 \times 100 \times 200$ & 5.36 & 0.82  & 9.1 & 4900 & 587 & 15  \\
\,\textit{RASJA, water, sides}& &  & & & & &   \\
  \hline
 Cazaubiel et al. 2021  \cite{Cazaubiel2021} \, & $11.5 \times 11.5 \times 9$ & 2  & 0.87 &  5 & 1000 & 122 &12  \\
  \,  \textit{Forcing by magnetic particles, water} &  &  \textit{18} &  &  &\textit{ 9000} & &\textit{ 60} \\
  \hline
Gorce \& Falcon 2022  \cite{Gorce2022} \, & $ 32 \times 32  \times 22$ & 1.6  & \textit{NA} &  5 & 650 & 100 &3  \\
  \,  \textit{Forcing by magnetic particles, water} &  &   &  &  & & & \\  
    \hline  
  \textbf{This work} & $40 \times 40 \times 66$ & 8.25 & 1.01 & 7.0 & 6603 & 461 & 33   \\
\,\textit{RASJA, free-surface} &     &  \textit{11} & \textit{0.89} & \textit{7.1} & \textit{8801} &\textit{558}   & \textit{71}   \\
\end{tabular}
\caption{{Properties of turbulent flow in the homogeneous and isotropic region,  for selected experiments with low mean flow. In this table, we compare the dimensions of the experiment (length $L$, width $W$ and height $H$), the horizontal velocity fluctuations $\tilde{\sigma}_u$, the anisotropy ratio $\tilde{\sigma}_w /  \tilde{\sigma}_u$, the longitudinal horizontal integral scale $L_{uu}$, the turbulent Reynolds number associated with the integral scale $Re_T$, the Taylor Reynolds number $Re_\lambda$ and the dissipation rate $\epsilon$. We consider in chronological order, two vertically oscillating grid experiments in water with free-surface (Brumley \& Jirka~\cite{BrumleyJFM1987} and Herlina \& Jirka~\cite{HerlinaJFM2008})~\footnote{{The turbulent Reynolds numbers reported by these authors is here divided by two accordingly to our definition of the turbulent Reynolds number.}}, one RASJA experiment in air (Carter et al.~\cite{CarterEF2016,CarterJFM2017}), three RASJA experiments in water (Variano \& Cowen~\cite{VarianoJFM2008}, Johnson \& Cowen~\cite{JohnsonJFM2017} and Esteban et al.~\cite{EstebanJFM2019}) and two experiments where the turbulence is driven by magnetic particles~\cite{Cazaubiel2021,Gorce2022}. Flow characteristic for some other RASJA experiments can be found in the recent review~\cite{Nezami2023}.  In this table, the typical values for high turbulence Reynolds are given in standard font. For Johnson \& Cowen~\cite{JohnsonJFM2017}, the values for a second configuration of jets are given in italics. For Cazaubiel et al. the largest accessible values characterized by LDV are given in italics. For our work, we give the values for experiment A in standard font and the values at the highest Reynolds number in italics.}}
\label{tab:compturb}
\end{ruledtabular}
\end{table}

{The level of turbulence reached within our experimental device is quantified by computing two common Reynolds numbers of the flow: the turbulent Reynolds number $\mathrm{Re}_T=\mathcal{U} L/\nu$} associated with the integral length scale and the {large-scale} velocity fluctuations,  and the Taylor-scale Reynolds number $\mathrm{Re}_\lambda=\mathcal{U}^2\sqrt{15/(\nu\epsilon)}$ linked to the typical size of velocity gradients and characterizing the turbulent cascade~\cite{Pope,Bailly}. {For our range of flow rates, we obtain  $\mathrm{Re}_T$ between {$991$ and $8801$} and  $\mathrm{Re}_\lambda$ between $148$ and $558$. We find {$20\, \mathrm{Re}_T/\mathrm{Re}_\lambda^2 \in [0.53;0.90]$}, which is compatible with theory predicting  $\mathrm{Re}_T\sim \mathrm{Re}_\lambda^2/20$~\cite{Pope}.
 The variables estimated in the previous subsections are then used to compute the corresponding values of these two Reynolds numbers, to compare the performances of our experimental device with other RASJA, grid-stirred or magnetically forced experiments in the Table \ref{tab:compturb}.}

To conclude this section, we demonstrate that our original experimental device associating a large pump with randomly opened jets constitutes a convenient facility to study homogeneous and isotropic turbulence with low mean flow and circulation and with a sufficient Taylor-scale Reynolds number to obtain a developed inertial range. The temporal velocity power spectrum in Fig.~\ref{fig:spectre_temporel_u_w} thus displays a power-law in $\omega^{-5/3}$ on more than one decade and the ratio of the longitudinal integral length to the Kolmogorov scale reaches values in the order of $500$. Moreover, by increasing the flow rate of the central pump, the turbulent Reynolds number can be varied, while keeping the longitudinal integral length around $8\,$cm, with good levels of isotropy and homogeneity and with low mean flow. 
Finally, we note that our device has been explored only for a specific configuration of the jet patterns, opening duration, and water-filling depth. Numerous flow configurations could be tested and used in further research in fundamental or applied turbulence.

\section{Turbulence anisotropy near the free surface}
\label{sec:nearFS}

In this section, we study the turbulent flow in the region where the free surface makes turbulence anisotropic, i.e., for $-6<z<0$\,cm. As reported in previous studies~\cite{BrumleyJFM1987,VarianoJFM2008}, the vertical size of this domain is of order the longitudinal integral scale. While the measurement of the free-surface deformation and the subsequent analysis will be reported in further dedicated work, we must note that the rms amplitude of the free-surface deformation $\sigma_\eta$ does not exceed $2\,$mm {for the range of turbulence level studied here}~\cite{JaminPhd2016}. By assimilating the integral scale {$L=\widetilde{L}_{uu} \approx 8$\,cm} as the size of the free-surface excitation by the flow, the ranges of the Froude number Fr$=\mathcal{U} /\sqrt{2\,g\,L}$ and Weber number We$= (\rho\,\mathcal{U}^2 \, L)/(2\,\gamma)$, built on the integral scale~\cite{BrocchiniJFM2001}, are respectively {$[0.010;\,0.088]$ and $[0.088;\,6.9]$}, {where} $g=9.81$ m~s$^{-2}$ is the gravity acceleration, $\rho=998$\,kg~m$^{-3}$ the water density and $\gamma \approx 0.07$ N~m$^{-1}$ the air-water surface tension. According to the qualitative analysis of free-surface deformations in the presence of turbulence by Brocchini and Peregrine~\cite{BrocchiniJFM2001}, given the intensity of turbulence {($\mathcal{U} \in [1.24; 11.0]\,$cm~s$^{-1}$)} with $L\approx8\,$cm, our experiments belong to the "flat" or the "wavy" domain. For these parameters, the "wavy" domain indeed starts at $\mathcal{U} \approx2.8\,$cm~s$^{-1}$ and the breaking transition occurs for $\mathcal{U}\approx 32\,$cm~s$^{-1}$~\cite{BrocchiniJFM2001}. Consequently, a breaking of the free surface is not expected and not observed for these levels of turbulence. As the deformations remain small compared to the integral length $L$, the free-surface dynamics is expected to act as a free-slip rigid surface in first approximation~\cite{BrocchiniJFM2001}. We note, that in the case of significant surface deformations, the velocity field characterization using PIV would require the simultaneous measurement of the free surface shape to apply the PIV algorithms only in the liquid domain. Such a challenging experimental implementation has not yet been performed for a free surface deformed by the turbulence and is limited to two-dimensional~\cite{SanchisExpFlu2011} or axisymmetric~\cite{jamin2015experiments} flows and interface deformations.

\subsection{Velocity fluctuations near the free surface}
\label{Velflucnear}
\begin{figure}
\centering
\includegraphics[width=11cm]{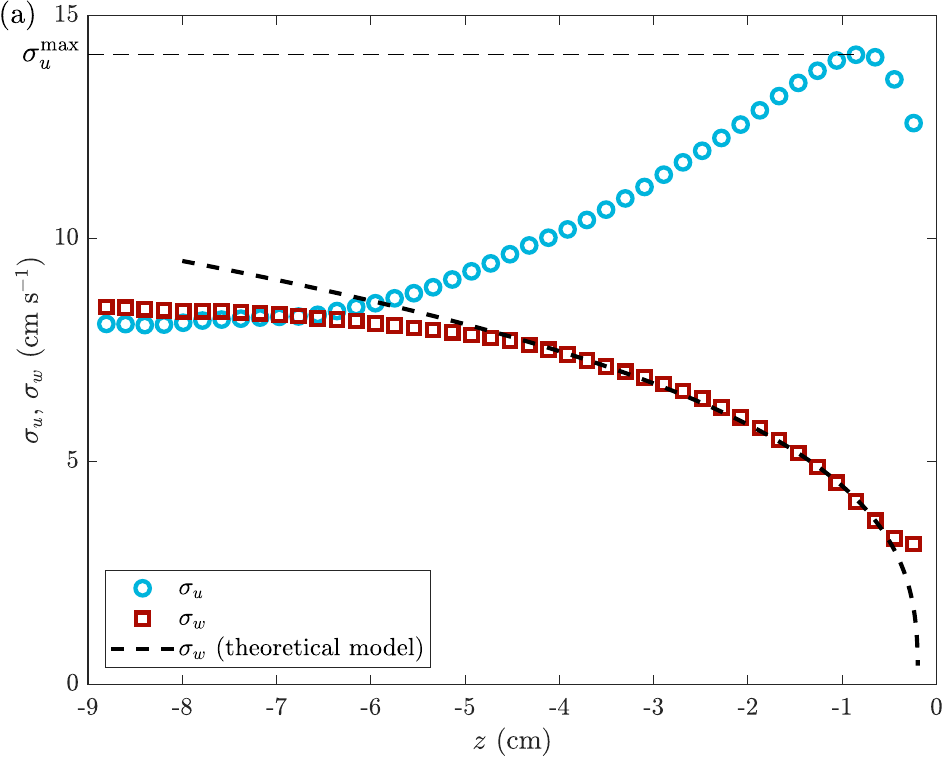}
\hfill
\includegraphics[width=5.65cm]{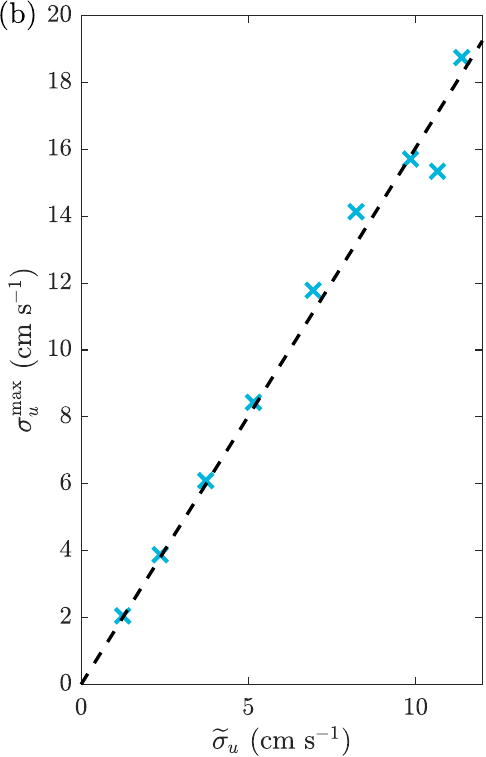}

\caption{ (a) Vertical profiles of {standard deviations for horizontal and vertical velocities $\sigma_u$ and $\sigma_w$}, averaged horizontally over $-5 < x < +5\,$cm. Theoretical model (dashed line) corresponds to Eq.\,\eqref{eq:sigmawmod} with $L=8\,$cm, $\beta=2.2$. {The evolution of vertical fluctuations as a function of the distance to the free surface $z$ is well described by this model for $z<0.6\,L \approx 4.8$  cm.} Parameters of Experiment A. (b) Maximum of $\sigma_u$ {from (a)} as a function of its value in the homogeneous and isotropic region, $\widetilde{\sigma}_u$. Dashed line corresponds to a linear regression: $\sigma_{u}^{\mathrm{max}}/\widetilde{\sigma}_{u}=1.60$ (95\,\% CI: [1.52; 1.68]).}
\label{fig:profil_spatial_vertical_vitesses}
\end{figure}

We use the PIV measurements to investigate the influence of the free surface on the turbulent velocity field for $-6<z<0$\,cm. In Figs~\ref{fig:sigma_u2Ddeform}{\,(a) and (b)}, the influence of the top free surface becomes clearly visible for $z>-4\,$cm. We note that the vertical fluctuations $\sigma_w$ show a good level of homogeneity along the horizontal coordinate $x$, whereas $\sigma_u$ is maximal at the tank center and significantly decreases on the sides of the images as visible in Fig.\,\ref{fig:profil_spatial_horizontal}\,(b). This behavior is likely due to the finite-size effects of the tank as stated by Variano and Cowen~\cite{VarianoJFM2008}.  The main effect of the free surface is displayed by the vertical dependency of velocity fluctuations. The strong modification of velocity fluctuations shown in Figs~\ref{fig:sigma_u2Ddeform}{\,(a) and (b)} is highlighted when averaging horizontally the vertical profiles of velocity as plotted in Fig.~\ref{fig:profil_spatial_vertical_vitesses}\,(a): $\sigma_w$ is found to decrease close to the free surface while $\sigma_u$ increases except over a 1-cm-thick layer just beneath the free surface. While we observe good isotropy for $z<-6\,$cm, the turbulent flow thus becomes strongly anisotropic near the free surface ($z=0$). This increase of the flow anisotropy when approaching the free surface is also reported in the experimental work of Variano and Cowen~\cite{VarianoJFM2008} and in the experiments performed with grid stirred tanks~\cite{BrumleyJFM1987,chiapponi2012experimental}. {In these studies like ours}, the turbulent flow becomes affected by the free surface for depth of order or smaller than the integral length $L$.

The increase of turbulence anisotropy when approaching a wall has been addressed theoretically first by Hunt and Graham~\cite{HuntJFM1978,HuntJFM1984}. Using the rapid distortion theory (RDT), they consider the displacement of a turbulent structure from the homogeneous isotropic region towards the interface and compute as a potential flow the modification of the structure in a linear framework without dissipation. The free surface plays the role of a rigid wall limiting vertical motions of the fluid. Then a transfer from vertical kinetic energy to horizontal kinetic energy occurs. The affected depth depends on the typical size of the largest eddies {$L$}. Indeed, Hunt \cite{HuntJFM1984} predicts for shear-free turbulence that the vertical velocity fluctuations near a {horizontal} wall scale as $\sigma_w\propto |z|^{1/3}$ in a layer defined by $-L\lesssim z \leq 0$, except in a viscous boundary layer close to the wall. Using numerical simulations, Calmet and Magnaudet investigate specifically the case of a free surface modeled as a free-slip rigid wall~\cite{CalmetMagnaudetJFM2003}. They find a thickness of the viscous layer of order $\delta_\nu\approx2.0\,L/\sqrt{2\,\mathrm{Re}_T}$ in agreement with previous experimental measurements~\cite{BrumleyJFM1987}. This formula gives $\delta_\nu\in[1.7; 5.1]\,$mm for our experiments where the largest values correspond to the lowest turbulent intensities. By taking into account this viscous boundary layer, they predict the following evolution of the vertical velocity fluctuations for $-L\lesssim z  \leq \delta_\nu$:

\begin{equation}
\sigma_w=\sqrt{\beta}\epsilon^{1/3}\left(\frac{L(|z|-\delta_\nu)}{L-\delta_\nu}\right)^{1/3}
\label{eq:sigmawmod}
\end{equation}
\noindent where the prefactor $\beta$ is theoretically computed to be $\beta= 1.784$~\cite{MagnaudetJFM2003}.
In stationary regime, this expression of $\sigma_w$ has been verified {for $-0.7\, L\lesssim z \lesssim \delta_\nu$} in numerical simulations~\cite{CalmetMagnaudetJFM2003} by choosing $\beta\approx 2.0$. Variano and Cowen \cite{VarianoJFM2008} experimentally measured $\beta \approx 1.5$ and observed that the model is valid for $z>-L$, the viscous layer being unresolved.

In Fig.\,\ref{fig:profil_spatial_vertical_vitesses}\,(a), we add the curve corresponding to Eq.\,\eqref{eq:sigmawmod}, which fits well with the experimental profile of vertical velocity fluctuations $\sigma_w$ for $-0.6\,L\leq z\leq0$ when choosing $\beta=2.2$ and the parameters $L$, $\epsilon$ and  $\mathrm{Re}_T$ characterizing the turbulence in the bulk, determined in Sect.~\ref{Homogeneousturbulence}. For the other flow rates (not shown), we find again a good agreement between the theoretical vertical fluctuation profile and experimental profile in the domain $-0.6\,L \leq z \leq 0$ for our range of  $\mathrm{Re}_T$ by choosing {$2.0 \lesssim \beta\lesssim 2.3$}.

Notice that the decrease of $\sigma_w$ slows down very close to the free surface [for $z>-0.5\,$cm in Fig.~\ref{fig:profil_spatial_vertical_vitesses}\,(a)]. This departure from the theoretical curve can be detected only for an intense enough turbulence ($\mathcal{U}\gtrsim 5$\,cm s$^{-1}$ or  $\mathrm{Re}_T\gtrsim 3000$). Moreover, the depth of this inflection point increases with forcing while $\delta_\nu$ decreases, implying that this observation is not related to the viscous sublayer. This effect is probably due to the small deformations (few millimeters) of the free surface caused by turbulence. Indeed, this effect is neither visible for the lowest forcings of our experiment nor in Variano \textit{et al.} experiment which displays only gentle free-surface deformation and where $\sigma_w$ seems to decrease to zero at the free surface. By taking into account the free-surface displacement $\eta(t)$ [i.e., $w(z=0,t)\approx\partial\eta/\partial t$], the free surface does not behave anymore as a rigid shear-free wall. {Consistently with our measurements,  Guo and Shen observe in their numerical simulations a nonvanishing value of $w$ in $z=0$ which increases with Froude number and thus with surface defomations~\cite{GuoJFM2010}. Measurements with better resolution near the free surface are needed to confirm and analyze this observation.

A theoretical prediction similar to Eq.\,\eqref{eq:sigmawmod} exists for $\sigma_u$ \cite{HuntJFM1984,MagnaudetJFM2003} 
but was shown to be valid only for ${-0.1L}<z<0$ \cite{CalmetMagnaudetJFM2003}, which is a too small range to be verified in our experimental setup. But the same studies also predicts {in first approximation} that $\sigma_u$ should monotonically increase when approaching the free surface and reach $\sigma_u^2(z=0)=1.5\,\widetilde{\sigma}_u^2$, i.e., an increase of $22\,\%$ for $\sigma_u$. As visible in Fig.\,\ref{fig:profil_spatial_vertical_vitesses}\,(a), we do not observe such a monotonic increase in our experiments as $\sigma_u$ reaches a maximum at $z\approx -0.9\,$cm. The position of this maximum appears to be independent of the forcing (not shown here), contrary to the size $\delta_\nu$ of the viscous sublayer which is thus not involved. The same phenomenon at a similar vertical location appears in Variano and Cowen \cite{VarianoJFM2008} experiment. This non-monotonic evolution of $\sigma_u$ is interpreted by these authors} as a contamination of the air-water interface by surfactants, which changes the boundary condition. In the case of a solid wall, $\sigma_u$ is expected to reach 0 in $z=0$ due to the no-slip boundary condition. For a free surface, one expects $du/dz|_{z=0}=0$ due to the absence of tangential forces, as observed in numerical simulations. However, a free surface contaminated with surfactants may have a behavior close to a solid wall as surfactants induce horizontal surface tension gradients creating tangential forces called Marangoni stresses~\cite{Lucassen1968}. {Such pollution of the free surface is very difficult to avoid in experiments using water~\cite{van1966boundary,miles1967surface,henderson1990single,hammack1993resonant,FrewJGR2004}. The influence of surfactants on the mechanical properties of a  free surface has been investigated for various levels of contamination in numerical simulations~\cite{ShenJFM2004,WissinkJFM2017}, showing a decrease of the value of $\sigma_u$ at the free surface, but without a non-monotonic evolution.

Nevertheless, the increase of $\sigma_u$ with $z$ before the maximum is consistent with the conversion of vertical fluctuations into horizontal ones. Then, we plot in Fig.\,\ref{fig:profil_spatial_vertical_vitesses}\,(b) the maximum of horizontal velocity fluctuations, $\sigma_u^{max}$, as a function of $\widetilde{\sigma}_u${, the value in the homogeneous and isotropic region}. We observe a clear linear relation {$\sigma_{u}^{\mathrm{max}}=1.6\,\widetilde{\sigma}_{u}$. This $60\,\%$} increase is much larger than the predicted $22\,\%$ by the RDT~\cite{HuntJFM1978,MagnaudetJFM2003}.

 Conversely, we can read an increase of only 10\,\% in figures of Variano and Cowen \cite{VarianoJFM2008,VarianoJFM2013}.
Guo and Shen \cite{GuoJFM2010} showed in their numerical simulations that this increase greatly depends on surface behavior. It reaches $45\,\%$ for the weakest surface deformations ($\mathrm{Fr}=0.22$ using our definition) and seems to be around $25\,\%$ and $20\,\%$ for Fr$=0.32$ and $\mathrm{Fr}=0.63$ respectively. Notice that Froude numbers for our experiments are smaller ($[0.010;\,0.088]$) which may explain an even larger increase. Interpretation of this over-amplification relative to the RDT has been proposed by Hunt \cite{HuntJFM1984} and Kit \textit{et al.} \cite{KitJFM1997}, {including} nonlinear vortex stretching near the interface, but was discredited by Magnaudet \cite{MagnaudetJFM2003}. For our experiment, this result may also be linked to the important horizontal inhomogeneity observed for horizontal velocity fluctuations near the free surface in Fig.\,\ref{fig:profil_spatial_horizontal}\,(b), which could concentrate the energy in the horizontal center of the tank.

To conclude, our measurements show that the turbulent velocity fluctuations are affected by the top free surface, with an enhancement of the horizontal fluctuations and a decrease of the vertical ones. These observations complete previous experimental~\cite{BrumleyJFM1987,VarianoJFM2008} and numerical~\cite{CalmetMagnaudetJFM2003,GuoJFM2010} results of the literature. {They are} in qualitative agreement with the predictions of the RDT describing the geometric deformation of a turbulent flow at the vicinity of a rigid surface~\cite{HuntJFM1978,HuntJFM1984,MagnaudetJFM2003}. However, we observe that near the surface, horizontal velocity fluctuations increase much more than predicted theoretically.

\subsection{Velocity power spectra near the free surface}

The modification of the turbulence at the vicinity of the free surface is also well characterized by the velocity power spectra. In Sec.~\ref{sec:spectres_caract}, we show in Fig.~\ref{fig:spectre_spatial_u_w_-6_4cm} that the energy spectra of horizontal and vertical velocities computed far from the free surface are {in satisfying} agreement with the Kolmogorov 41 predictions for homogeneous and isotropic turbulence.
The anisotropy of turbulent fluctuations when {we get closer to the free surface}, mentioned in Sect.~\ref{Homogeneousturbulence}, is also reflected in the spatial spectra recorded close to the free surface, {as shown} in Fig.~\ref{fig:spectre_spatial_u_w_-1cm} for $z=-1.1\,$cm. At large scale (small horizontal wavenumber $k_x$), the amplitude of the horizontal velocity spectrum $S_{u,k_x}$ is enhanced, whereas the one of the vertical velocity spectrum $S_{w,k_x}$ is reduced. This observation is consistent with the conversion of vertical fluctuations into horizontal ones, discussed in Sect.~\ref{Velflucnear} for the profile of horizontal fluctuations.

\begin{figure}
\centering
\includegraphics[width=11cm]{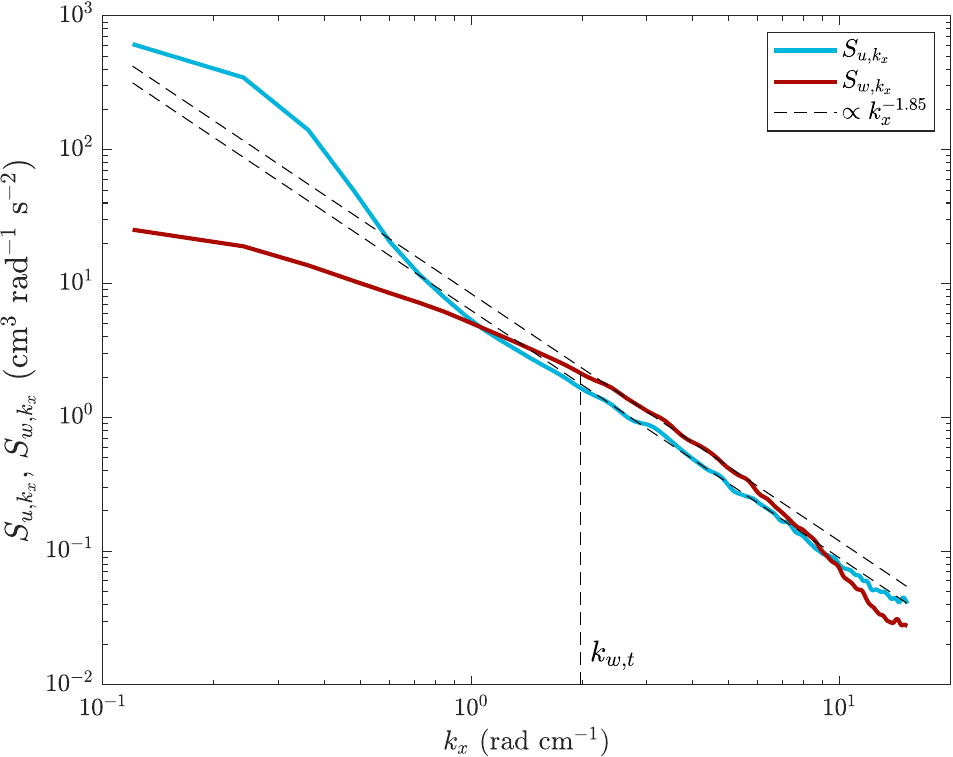}
\caption{Longitudinal (light blue) and transverse (dark red) spatial power spectra of horizontal and vertical turbulent velocity fluctuations $S_{u,k_x}$ and $S_{w,k_x}$ along $x$ as functions of horizontal wavenumber $k_x$ close to the free surface and averaged over time. Dashed lines represent the same power law in $k_x^{-1.85}$ and the same prefactor as in Fig.\,\ref{fig:spectre_spatial_u_w_-6_4cm}. Parameters of Experiment A; $y=0$ and $z=-1.1\,$cm.}
\label{fig:spectre_spatial_u_w_-1cm}
\end{figure}

To have a better grasp of what happens at different scales, we display in Fig.\,\ref{fig:spectre_spatial_u_w_multi}, the (a) longitudinal and (b) transverse spatial power spectra at different depths. A transition wavenumber $k_{u,t} \approx 2\pi/L$, roughly independent of the depth, appears to separate the longitudinal spectrum into two parts. At small scales ($k_x > k_{u,t}$), we observe the turbulence remains isotropic and satisfyingly follows the same power law in $k^{-1.85}$ with the same parameters $C_\parallel'$ and $\epsilon$ as in the homogeneous and isotropic region.
At large scales ($k_x < k_{u,t}$), the amplitude of the longitudinal spectrum in Fig.~\ref{fig:spectre_spatial_u_w_multi}\,(a) increases when approaching the free surface. {Moreover, it should be noted that in our experiment, due to the finite size of our setup, the large-scale part of the horizontal spatial spectrum is affected by the significant horizontal inhomogeneity observed for horizontal velocity near the free surface in particular on the edges as illustrated in Fig.\,\ref{fig:profil_spatial_horizontal}\,(b). This induces an artificially enhanced longitudinal spectrum at large scales.}

\begin{figure}
\centering
\includegraphics[width=8.8cm]{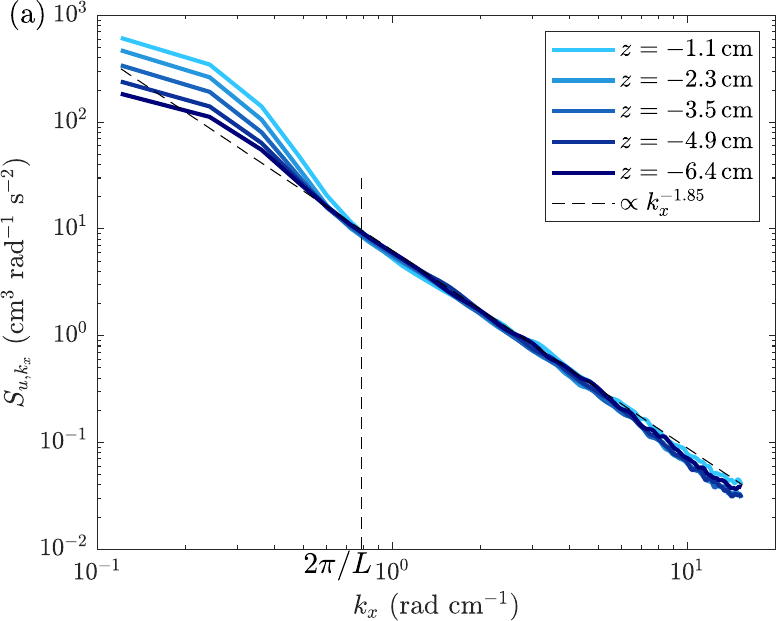}
\hfill
\includegraphics[width=8.8cm]{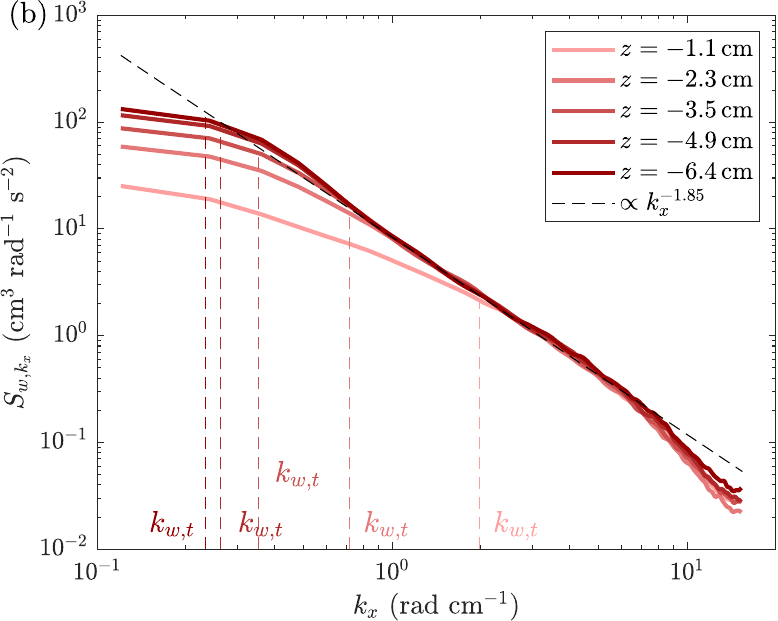}
\caption{(a) Longitudinal and (b) transverse spatial power spectra of horizontal and vertical turbulent velocity fluctuations $S_{u,k_x}$ and $S_{w,k_x}$ along $x$ as functions of horizontal wavenumber $k_x$ averaged over time for multiple depths and for $y=0$. Dashed lines represent the same power law in $k_x^{-1.85}$ and the same prefactor as in Fig.\,\ref{fig:spectre_spatial_u_w_-6_4cm}. Parameters of Experiment A.}
\label{fig:spectre_spatial_u_w_multi}
\end{figure}

For the transverse spectrum in Fig.\,\ref{fig:spectre_spatial_u_w_multi}\,(b), the transition between large scales (reduced spectrum amplitude) and small scales (unchanged spectrum) occurs for a wavenumber $k_{w,t}$ increasing when getting closer to the free surface. $k_{w, t}$ is defined as the wavenumber for which the large-scale part of $S_{w,k_x}$ deviates from the $k_x^{-1.85}$ power law and is below the corresponding dashed line [see Fig.\,\ref{fig:spectre_spatial_u_w_-1cm} and Fig.\,\ref{fig:spectre_spatial_u_w_multi}\,(b)].
We plot in Fig.\,\ref{fig:spectre_spatial_ratio_kwt}\,(a) the inverse of $k_{w,t}$ as a function of the depth. We observe that the transition is {roughly} linearly linked to the depth {for $z>-6$\,cm} with $|z|\,k_{w,t}\approx 1.4$\,rad.

\rev{The evolution of isotropy can also be studied thanks to Fig.\,\ref{fig:spectre_spatial_ratio_kwt}\,(b) where the ratio of spectra $S_{w,k_x}/S_{u,k_x}$ is plotted as a function of $k_x$. Indeed, the relationship between longitudinal and transverse spectra is expressed as $S_{w,k_x}=(S_{u,k_x}-k_xS_{u,k_x}')\,/\,2$ in homogeneous and isotropic turbulence~\cite{Pope}. Thus a theoretical ratio of $S_{w,k_x}/S_{u,k_x}=4/3\approx1.33$ (or 1.425 for a power law in $k_x^{-1.85}$) is expected in the inertial range. While we observe a plateau close to this value (the horizontal dashed line corresponds to $4/3$ in the figure and the plateau ) for $z=-6.4$\,cm (darkest line), the ratio progressively deviates from its theoretical value for low $k_x$ when getting closer to the free surface. It shows that anisotropy increases at large scale, while it is not the case at small scales.}

{These results are partially explained with the RDT. It states that for eddies larger than the depth $|z|$, vertical velocity fluctuations are converted into horizontal ones~\cite{HuntJFM1978}, and the transverse spectral energy density for vertical velocity decreases while the longitudinal one for horizontal velocity increases. Accordingly, this theory predicts that both horizontal and vertical spatial spectra $S_{u,k_x}$ and $S_{w,k_x}$ remain unchanged near the free surface for small scales, i.e., for $k_x \gtrsim 1/|z|$ ~\cite{HuntJFM1978,HuntJFM1984}. The same $-5/3$ power law is then expected at these scales, a consequence of the structure of the turbulence in the isotropic and homogeneous region far from the surface. For larger scale ($k_x \lesssim 1/|z|$), the amplitude of the longitudinal spectrum of horizontal velocities $S_{u,k_x}$ should be enhanced while $S_{w,k_x}$ is strongly reduced. The latter is supposed to reach a plateau value evolving in $z^{-5/3}$ for $k_x \ll 1/L$ whereas the plateau value of the former should stay the same.}

{In our measurements, we observe a clear enhancement of the amplitude of the longitudinal spectrum of horizontal velocity $S_{u,k_x}$ at the vicinity of the free surface below a transition wavenumber $k_{u,t} \approx 2\pi/L$ while the spectrum remains unchanged above $k_{u,t}$ [see Fig.\ref{fig:spectre_spatial_u_w_multi}\,(a)]. However, according to the RDT, the transition wavenumber $k_{u,t}$ should be in $1/|z|$, whereas it remains nearly constant in our experiments. Yet, the region $k_x \ll 1/L$ in which the RDT predicts an unchanged plateau is also not spatially resolved in our experiments. The enhancement of $S_{u,k_x}$ in the intermediate region ($k_x \lesssim 1/|z|$) is also expected according to the RDT but less markedly. Indeed, spectral power density is not supposed to be over $S_{u,k_x} (k_x = 0, z = -\infty)=2\widetilde{\sigma}_{u}^2L/\pi=360$\,cm$^3$~rad$^{-1}$~s$^{-2}$ for experiment A~\cite{HuntJFM1978,HuntJFM1984}, but we see in Fig.\,\ref{fig:spectre_spatial_u_w_multi}\,(a) that $S_{u,k_x}$ reaches values larger than 600\,cm$^3$~rad$^{-1}$~s$^{-2}$ close to the free surface.}

{Regarding the vertical velocity spectrum, the transition wavenumber follows $k_{w,t}\sim1/|z|$ [see Fig.\,\ref{fig:spectre_spatial_ratio_kwt}\,(a)]. For $k_x>k_{w,t}$ the spectrum remains unchanged and {conversely} the amplitude of the spectrum decreases. This observation is in agreement with the RDT and could be explained by a better homogeneity along $x$ for the vertical velocities than the horizontal ones close to the free surface [see Fig.\,\ref{fig:profil_spatial_horizontal}\,(b)].}

{The partial disagreement between our observations and the RDT could be indeed explained by the inhomogeneity of the flow, which may be caused by the finite size of the tank. However, we emphasize that the RDT has some limitations due to restrictive hypotheses, which are not fully met in experiments and environmental flows. The RDT describes indeed the linear deformation by the free surface of a “frozen” potential flow in a decaying regime without viscous dissipation. In our experiment, with continuous forcing, surface turbulence is continuously fed by the impingement of turbulent structures from below. In the absence of more appropriate models in the literature, we use the RDT for qualitative insights only.}

\begin{figure}
\centering
\includegraphics[width=8.8cm]{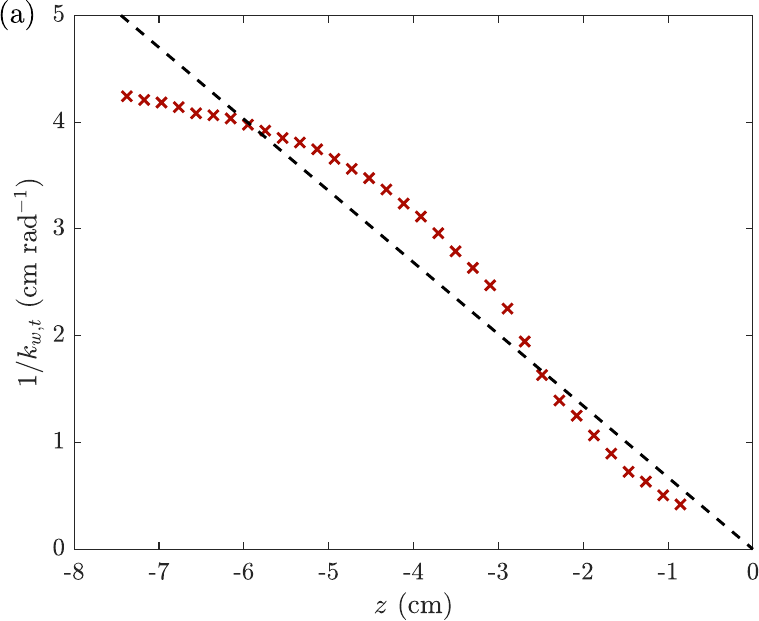}
\hfill
\includegraphics[width=8.8cm]{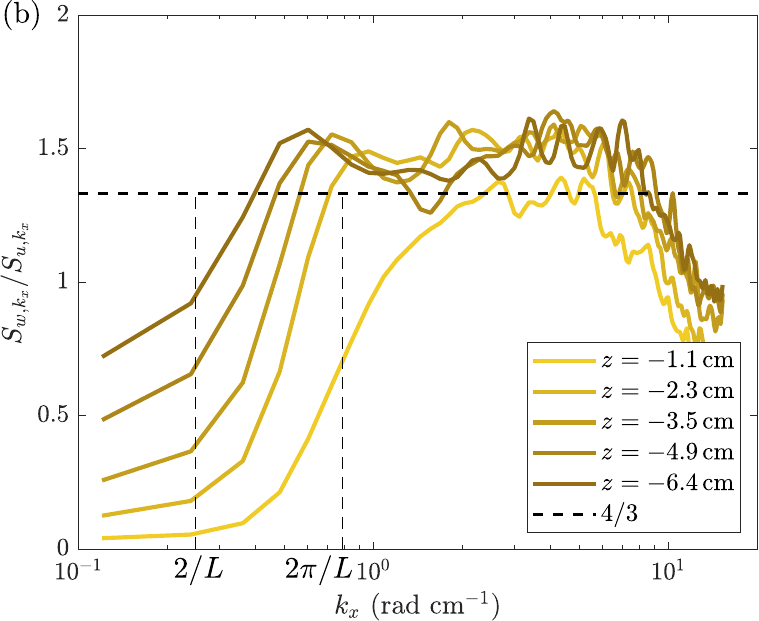}
\caption{(a) Inverse of the wavenumber $k_{w,t}$ for which $S_{w,k_x}$ deviates from $k_x^{-1.85}$ power law [see Figs~\ref{fig:spectre_spatial_u_w_-1cm} and \ref{fig:spectre_spatial_u_w_multi}\,(b)], as a function of depth. Dashed line is a linear fit corresponding to the equation $z\,k_{w,t}=-b$, with {$b=1.49$\,rad (95\,\% CI:  [1.43; 1.55]\,rad)}. (b) \rev{Ratio of the transverse spatial spectrum to the longitudinal one $S_{w,k_x}/S_{u,k_x}$ as a function of horizontal wavenumber $k_x$ averaged over time for multiple depths and for $y=0$. The horizontal dashed line represents the theoretical ratio $4/3$ in the inertial range for homogeneous and isotropic turbulence.} }
\label{fig:spectre_spatial_ratio_kwt}
\end{figure}

\begin{figure}
\centering
\includegraphics[width=11cm]{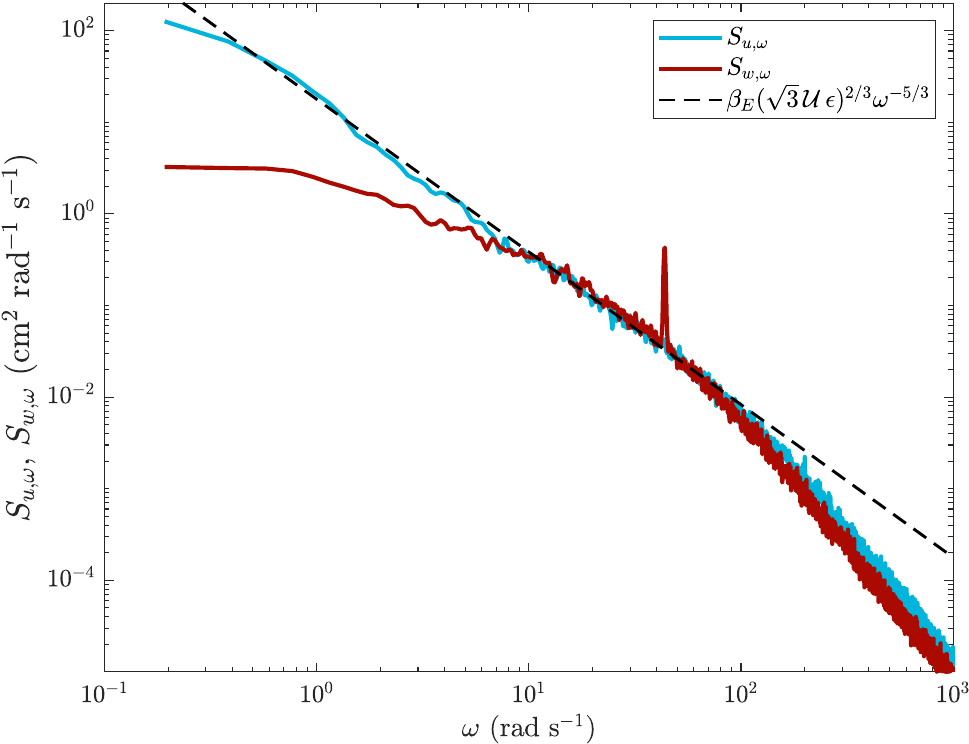}
\caption{Horizontal and vertical velocity temporal {power spectra} $S_{u,\omega}$ and $S_{w,\omega}$ as functions of the frequency $\omega$ for $x=0$, $y=-7$\,cm and depths $z=-1.0\,$cm and $z=-1.3\,$cm, for horizontal and vertical velocities $u$ and $w$ respectively. Dashed line represents the same theoretical law predicted by Tennekes \cite{TennekesJFM1975} choosing the same value $\beta_E=0.7$ as in Fig.\,\ref{fig:spectre_temporel_u_w}. Duration: 1200\,s. $Q_j=0.25\,$L s$^{-1}$.}
\label{fig:spectre_temporel_u_w_FS}
\end{figure}

{The comparison of our experimental results with the RDT detailed above is summarized in Table~\ref{tab:RDT_summary}.}

\begin{table}
   \centering
\begin{tabular}{|*{7}{c|}}
\hline
&
\multicolumn{3}{|c}{Longitudinal spectrum $S_{u,k_x}$} & \multicolumn{3}{|c|}{Transverse spectrum $S_{w,k_x}$} \\ \hline
& Transition & Small scales & Large scales & Transition & Small scales & Large scales \\
& wavenumber $k_{u,t}$ & $k_x>k_{u,t}$& $k_x<k_{u,t}$ & wavenumber $k_{w,t}$ & $k_x>k_{<,t}$ & $k_x<k_{w,t}$\\ \hline 
RDT & $\propto1/|z|$ & invariant & small increase & $\propto1/|z|$ & invariant & decrease\\ \hline
Exp. observations & $\approx 2\pi/L$ & invariant & large increase & $\propto1/|z|$ & invariant & decrease \\ \hline
Agreement & No & Yes & Partial & Yes & Yes & Yes\\ \hline

\end{tabular}
\caption{{Comparison of the predictions of the RDT (first line) with the experimental observations of the present paper (second line) regarding spatial power spectra $S_{u,k_x}$ and $S_{w,k_x}$.}}
\label{tab:RDT_summary}
\end{table}

We complete the PIV measurements, with temporal spectra acquired using LDV at $z \approx -1$\,cm and displayed in Fig.~\ref{fig:spectre_temporel_u_w_FS}. The dashed line corresponds to the Tennekes model in $\omega^{-5/3}$ with the same $\beta_E$, $\mathcal{U}$ and $\epsilon$ values as in Sect.~\ref{sec:spectres_caract} and Fig.\,\ref{fig:spectre_temporel_u_w}. Then, we observe the velocity fluctuations follow the same equation as in the homogeneous and isotropic region in the {mid-range} of frequencies. In addition, the $\omega^{-5/3}$ power law is considerably extended towards low frequencies ($\omega \lesssim 3\,$rad~s$^{-1}$) for the horizontal velocity spectrum $S_{u,\omega}$, whereas the large-scale vertical velocity spectrum $S_{w,\omega}$ is reduced. The fact that the horizontal temporal velocity spectrum scales as $\omega^{-5/3}$ even for low frequencies while it is not the case of the spatial spectrum in Fig.\,\ref{fig:spectre_spatial_u_w_multi}\,(a) is surprising as the temporal spectrum is supposed to result from spatial structure advection, according to Tennekes model~\cite{TennekesJFM1975}. Besides, the significant extension towards low frequencies of the Kolmogorov spectrum of horizontal velocity close to a free surface despite the flow anisotropy also constitutes another surprising observation, which has been pointed out recently by Flores et al.~\cite{FloresJFM2017}. {These authors using numerical simulations of decaying free-surface turbulence attribute this fact to an inhibition of vertical motions by analogy with the stratified turbulence~\cite{FloresJFM2017}. Contrary to the RDT, where the inertial range near the free surface is linked to the one observed in the homogeneous region, the analogy with stratified turbulence makes this $-5/3$ power law near the free surface specific to a layer $\vert z \vert \lesssim 0.1\,L$ and independent from the turbulent flow below. At the vicinity of the free surface, the cascade is controlled by the local dissipation rate which is enhanced by the strong vertical shearing of horizontal velocities. The vertical velocities are also neglected and thus the vertical spectrum is not addressed by this analogy. However, in our experiments, as illustrated in  Fig.~\ref{fig:spectre_temporel_u_w_FS}, the spectrum $S_{u,\omega}$ follows well the power law established in homogeneous region using the same dissipation rate $\epsilon$ and the same prefactors as in the bulk. This is in disagreement with the hypothesis of enhanced local dissipation rate in the analogy with the stratified turbulence.}

 {Finally, suppression at low wavenumbers or frequencies of the vertical velocity spectrum and extension of the power law spectrum for the horizontal component above the isotropic range have been reported only by a few experimental studies with oscillating grids~\cite{BrumleyJFM1987,HerlinaJFM2008}, random actuated jets~\cite{VarianoJFM2013}, or field measurements at a river surface~\cite{chickadel2011infrared}.}

\subsection{Integral length scales near the free surface}

We now study experimentally the behavior of integral length scales when approaching the free surface. The typical size of turbulent eddies is measured by computing, as the function of the vertical coordinate $z$, the longitudinal integral length $L_{uu}$ for the horizontal velocity and the transverse one $L_{ww}$ for the vertical velocity using the PIV measurements and the same methods as in Sect.~\ref{Homogeneousturbulence}.
We use normalized autocorrelation functions $\rho_{uu}$ and $\rho_{ww}$ as functions of longitudinal increment $r_x$ and of depth $z$ as defined in Sect.~\ref{sec:echelles} and Eq.\,\eqref{eq:aurx}. {As the turbulence is not isotropic close to the free surface, we estimate the integral lengths by exponential fits.}

\begin{figure}
\centering
\includegraphics[width=11cm]{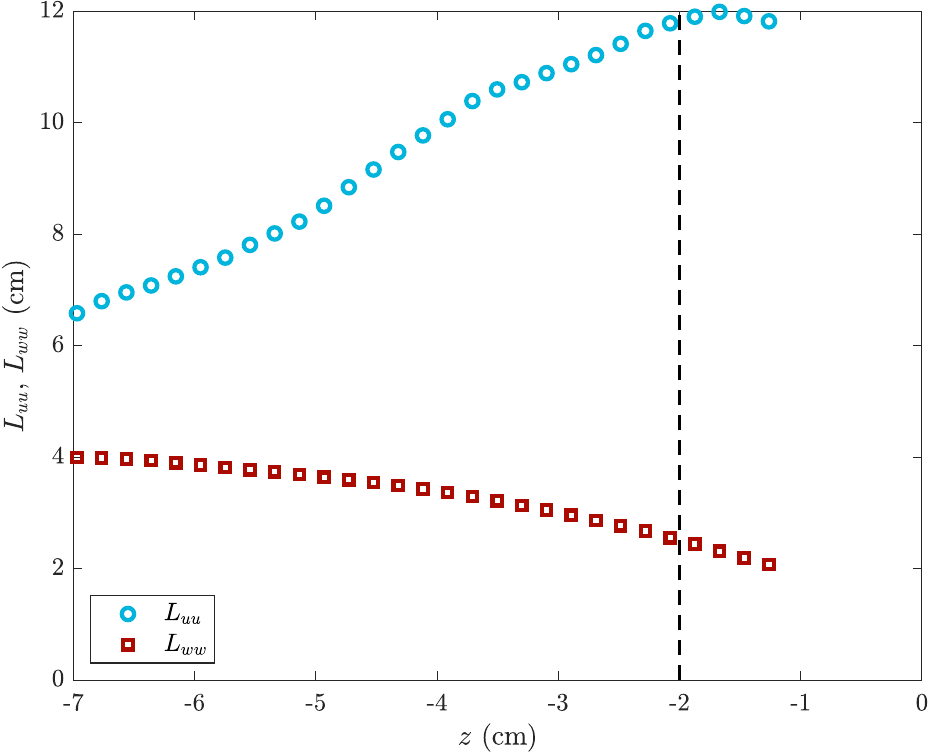}
\caption{Longitudinal ($L_{uu}$, circles) and transversal ($L_{ww}$, squares) integral length scales of horizontal and vertical velocities respectively as a function of depth $z$. Parameters of Experiment A. {$y=0$.}}
\label{fig:Lvsz}
\end{figure}

{However, these exponential fits  are quite approximate, especially for $L_{uu}$,} and we limit our discussion to a qualitative evolution of the integral scales of the turbulent flow when approaching the free surface. For $z>-2$\,cm, $\rho_{ww}$ displays some inconsistencies, probably due to the low values of $w$ and thus an increased sensitivity to artifacts. We plot in Fig.~\ref{fig:Lvsz} $L_{uu}$ and $L_{ww}$ for various depths. We observe that $L_{uu}$ increases near the free surface which is in strong disagreement with the decrease predicted by the RDT, where  $L_{uu}$ is predicted to slightly diminish to reach $L_{uu}(z=0) = 2/3\,\widetilde{L}_{uu}$ (due to the relation $L_{uu}=\widetilde{L}_{uu}\,\widetilde{\sigma_u}^2 / \sigma_u^2 $)~\cite{HuntJFM1978,HuntJFM1984}.  {This observation may be caused by the finite-size effects leading to horizontal heterogeneity of $u$ close to the free surface. }

Besides, $L_{ww}$ decreases experimentally when approaching the free surface. It is well predicted theoretically but with a lower rate than the predicted scaling $L_{ww} \approx 1.96\,|z|$~\cite{HuntJFM1984}, which is unsurprising as this prediction is valid for $z/L\ll 1$ only.
We notice that when moving away from the turbulence-producing area, $L_{uu}$ also increases in the homogeneous and isotropic region ($z<-6\,$cm). This observation was also reported in grid-stirred experiments~\cite{BrumleyJFM1987}. Further experimental, numerical, and theoretical studies are thus requested to understand the behavior of the local integral scale for turbulent flows in the vicinity of a free surface.

\section{Conclusion and perspectives}
\label{Conclusion}
We propose an evolution of the random actuated jet array (RASJA) device, which is now commonly used to produce homogeneous and isotropic turbulence with a small mean flow\red{~\cite{Nezami2023}}. Here, instead {of using an independent bilge pump for each jet}, the jets are connected to a central powerful pump, whereas the jets can be independently turned on and off. At a given time 16 jets over 64 fire simultaneously with a typical opening time of $3$\,s and a random choice of the position and of the opening duration. Using this protocol in a glass tank of dimensions $40 \times 40 \times 75$\,cm and filled up to a water depth of $66\,$cm, we characterize the properties of the turbulent flow mainly using PIV measurements in a vertical plane below the free surface. For a fluid depth larger than the integral length and sufficiently far from the jets, i.e., in the range $-9 <z <-6$\,cm, we recover the properties of homogeneous and isotropic turbulence, whose intensity can be adjusted over one order of magnitude by changing the flow rate of the central pump. {
We} reach at the maximum a turbulent Reynolds number {$\mathrm{Re}_T=8801$}, a Taylor Reynolds number {$\mathrm{Re}_\lambda=558$} and a dissipation rate {$\epsilon=71\,$cm$^2$ s$^{-3}$}. The turbulent fluctuations increase linearly with the flow rate per jet, vary between $1$ and $11$\,cm s$^{-1}$ and remain at least three times larger than the local values of the mean flow. Moreover, we observe that the integral length scale $L$ (assimilated to the longitudinal one) is not significantly affected by the forcing amplitude and remains close to {$8\,$cm}, a length which seems imposed by the geometry of the jet array and of the water tank. Therefore, the turbulent intensity can be varied without changing too much the large-scale structure. Our experimental device thus proposes an efficient facility to study fundamental homogeneous isotropic turbulence without a large mean flow in the laboratory. Turbulent transport of particles or multiphasic flows could also be investigated.

{Our study aims to characterize how turbulence is modified by an air-water top-free surface.} Despite the ubiquity of this situation in nature, this problem has been investigated in few experiments, mainly to quantify the gas exchanges with the atmosphere. 
With our experimental setup, we observe the modification of the turbulent flow near the free surface on a depth of order the integral length, i.e., in the domain $-6 < z < 0$\,cm. The deformations of the free surface by the subsurface {turbulence remain} sufficiently small in first approximation (below $2\,$mm) to treat the free surface as a rigid flat wall. The turbulence becomes strongly anisotropic when approaching the free surface, with a decay of the vertical velocity fluctuations and an enhancement of horizontal ones. The application of the rapid distortion theory (RDT) for shear-free turbulence near a wall initially proposed by Hunt and Graham~\cite{HuntJFM1978,HuntJFM1984} and improved by Magnaudet and Calmet~\cite{CalmetMagnaudetJFM2003,MagnaudetJFM2003} describes satisfyingly the experimental profile of the vertical velocity fluctuations using the turbulent parameters measured in the homogeneous isotropic region. We also report a predicted increase of the horizontal velocity fluctuations but reaching a maximum located about 1\,cm below the free surface, not predicted theoretically. The horizontal velocity increase is also larger than the one predicted by the RDT. This observation, also reported in some previous numerical simulations~\cite{GuoJFM2010}, does not have to date a satisfying theoretical explanation. In our case, this may be explained by finite size effects, observed near the free surface and leading to a concentration of horizontal velocity fluctuations in the horizontal center of the tank.

The modification of velocity power spectra $S_{u,k_x}$ and $S_{w,k_x}$ due to the free surface have also been investigated. At small scales, despite the anisotropy and the inhomogeneity of the flow, the velocity power spectra satisfyingly follow a power law in $k^{-1.85}$ like for homogeneous and isotropic turbulence {and with the same parameters}, suggesting that the small turbulent eddies remain unaffected by the free surface. Consistently with the decay of vertical velocity fluctuations for increasing $z$, the vertical velocity power spectrum $S_{w,k_x}$ displays also a smaller value at large scales, in agreement with the predictions of the RDT. The transition between large-scale and small-scale behaviors occurs for a transition wavenumber $k_{w,t}\sim1/|z|$ which is also predicted. In contrast, the horizontal velocity power spectrum $S_{u,k_x}$ is amplified at large scales, i.e., $k_x\lesssim2\pi/L$. The RDT predicts this increase but with a lesser amplitude and the predicted transition between {large-scale} and {small-scale} behaviors is predicted to occur at a wavenumber $k_{u,t}\sim1/|z|$. The temporal spectrum $S_{u,\omega}$ measured locally by LDV shows that the turbulent cascade in {$\omega^{-5/3}$} spreads at larger scales, as pointed out by Flores et al.~\cite{FloresJFM2017}.

Finally, we show experimentally a clear increase of the horizontal longitudinal integral scale when approaching the free surface, as well as a decrease {in} the vertical transverse integral scale. The former observation is in contradiction with the RDT, which predicts a decrease {in} the horizontal longitudinal integral scale.

To {sum up, our experimental findings} confirm predictions from the RDT regarding vertical velocity fluctuations when approaching the free surface: the quantitative evolution of $\sigma_w$, the decrease of the power spectrum $S_{w,k_x}$ at large scales and its constancy at small scales, as well as the transition wavenumber $k_{w,t}\sim1/|z|$ between these two parts, and the qualitative decrease of the transverse vertical integral scale. Regarding horizontal velocity fluctuations, we have discrepancies between our experimental data and RDT, quantitatively for some aspects (evolution of $\sigma_u$ and of large scale spatial power spectrum), but even qualitatively (evolution of transition wavenumber $k_{u,t}$ and of longitudinal horizontal integral scale).

We explain these differences mainly by finite-size effects, the boundaries concentrating horizontal velocity energy in the center of the tank.
Besides, the RDT assumes the linear deformation of a ``frozen'' potential flow due to geometrical obstacles and neglects any viscous effects, while the impingement of turbulent structures from below feeds continuously the surface turbulence and we may expect nonlinearities. Indeed, the emergence of intense turbulence and {large}-scale circulations appears as a consequence of the strong anisotropy leading to nearly two-dimensional flows. The free surface is also modeled as a stress-free interface but with no deformation. Therefore, {RDT} cannot accurately describe the turbulence close to a free surface, {but} is the only theoretical work applicable to our measurements to our knowledge. Further experimental, numerical, and theoretical work would thus be helpful to {better} understand the {influence} of a free surface on hydrodynamic turbulence.

\begin{acknowledgments}
We gratefully thank  Alexandre Lantheaume, Yann Le Goas, David Charalampous, Arnaud Grados, Alexandre Di Palma, Laurent R\'ea, Stefano Lun Kwong and Mathieu Receveur for technical assistance and help in the conception and building of the experimental device. We acknowledge the Sphynx group at CEA, France for lending a LDV apparatus for the first set of experiments. We are also indebted to Evan Variano (UC Berkeley) for {advice} and discussions about RASJA experiments. This work was funded by the French National Research Agency (ANR TURBULON project No. ANR-12-BS04-0005 and ANR DYSTURB project No. ANR-17-CE30-0004). T. Jamin was partially supported by the DGA (Direction Generale de l'Armement, France) and the CNRS.
\end{acknowledgments}

\end{document}